\documentclass[journal, english, twoscolumn, journal]{IEEEtran}

\usepackage{cite}
\usepackage{graphicx}
\usepackage{amssymb, amsmath, amsthm}
\usepackage{amsfonts}
\usepackage{multirow}
\usepackage{bm}
\usepackage{color}
\usepackage{algorithm}
\usepackage{algorithmic}
\usepackage{amsmath} 
\usepackage{xcolor}
\usepackage{flushend}
\usepackage{setspace}
\usepackage{placeins}
\usepackage{booktabs}
\usepackage{lipsum}
\usepackage{stfloats}
\usepackage{subfigure}
\usepackage{longtable}
\usepackage{makecell}
\usepackage{url}
\newtheorem{remark}{Remark}

\newtheorem{theorem}{Theorem}

\begin{document}
\title{UGV-assisted Wireless Powered Backscatter Communications for Large-Scale IoT Networks}
\author{Erhu Chen, Peiran Wu, \IEEEmembership{Member,~IEEE}, Yik-Chung Wu, \IEEEmembership{Senior Member,~IEEE}, \\ and Minghua Xia, \IEEEmembership{Senior Member,~IEEE}
\thanks{Manuscript received February 09, 2021; revised July 20, 2021; accepted October 01, 2021. This work was supported by the National Natural Science Foundation of China under Grants 62171486, 61801526, and U2001213. (\textit{Corresponding author: Minghua Xia.})}	
\thanks{Erhu Chen, Peiran Wu and Minghua Xia are with the School of Electronics and Information Technology, Sun Yat-sen University, Guangzhou 510006, China (e-mail: chenerh@mail2.sysu.edu.cn, \{wupr3, xiamingh\}@mail.sysu.edu.cn).}
\thanks{Yik-Chung Wu is with the Department of Electrical and Electronic Engineering, The University of Hong Kong, Hong Kong (e-mail: ycwu@eee.hku.hk).}
\thanks{Color versions of one or more of the figures in this article are available online at https://ieeexplore.ieee.org.}
\thanks{Digital Object Identifier XXX}
}

\markboth{IEEE Transactions on Wireless Communications, accepted for publication}
{Chen \MakeLowercase{\textit{et al.}}: UGV-assisted Wireless Powered Backscatter Communications for Large-Scale IoT Networkss}

\maketitle

\IEEEpubid{\begin{minipage}{\textwidth} \ \\[12pt] \centering  \copyright\ 2021 IEEE. Personal use is permitted, but republication/redistribution requires IEEE permission. \\
See \url{https://www.ieee.org/publications/rights/index.html} for more information.\end{minipage}}

\begin{abstract}
\noindent Wireless powered backscatter communications (WPBC) is capable of implementing ultra-low-power communication, thus promising in the Internet of Things (IoT) networks. In practice, however, it is challenging to apply WPBC in large-scale IoT networks because of its short communication range. To address this challenge, this paper exploits an unmanned ground vehicle (UGV) to assist WPBC in large-scale IoT networks. In particular, we investigate the joint design of network planning and dynamic resource allocation of the access point (AP), tag reader, and UGV to minimize the total energy consumption. Also, the AP can operate in either half-duplex (HD) or full-duplex (FD) multiplexing mode. Under HD mode, the optimal cell radius is derived and the optimal power allocation and transmit/receive beamforming are obtained in closed form. Under FD mode, the optimal resource allocation, as well as two suboptimal ones with low computational complexity, is developed. Simulation results disclose that dynamic power allocation at the tag reader rather than at the AP dominates the network energy efficiency while the AP operating in FD mode outperforms that in HD mode concerning energy efficiency.
 \end{abstract}

\IEEEpubidadjcol
\section{Introduction} \label{Introduction}

\IEEEPARstart{W}{ireless} powered backscatter communications (WPBC) have been widely adopted in ultra-low-power communication, such as radio-frequency identification (RFID) tags and Internet of Things (IoT) \cite{duan2020ambient}. In general, a WPBC system consists of three components: a carrier emitter, a tag reader, and multiple backscatter tags. At first, the carrier emitter generates radio frequency (RF) signals to activate backscatter tags. Then, each tag harvests the energy of incident RF signals for its internal circuit operations and modulates its information bits by simply reflecting the received carrier signals with different antenna impedances. Finally, the tag reader receives the backscattered signals and decodes them. Thanks to the low power consumption of tags, WPBC enables operation in a battery-free manner. Moreover, instead of generating RF signals by a tag itself, a backscatter tag does not require conventional RF components like local oscillators, mixers, and converters, thereby significantly facilitating the implementation in IoT.
\IEEEpubidadjcol

\subsection{Related Works and Motivation}
Despite the idea of backscatter communications was conceived by Stockman as early as in 1948 \cite{Stockman1948}, it is not put into practice until recent years. In 2012, a practical backscatter receiver equipped with a non-linear near-optimal detector was designed in \cite{kimionis2012bistatic}, and subsequently the same authors proposed to deploy multiple carrier emitters to increase the coverage of wireless sensor networks \cite{kimionis2014increased}. Furthermore, a backscatter network was designed in \cite{liu2017full} to enable both one-way wireless power transfer (WPT) and two-way wireless information transfer. These works are devoted to WPBC applications in either point-to-point scenario or small-scale networks. 

The communication range of WPBC is very limited due to the small amount of energy harvested by a backscatter tag \cite{MXia2015TSP}. To extend the coverage of WPBC, dedicated power beacons can be employed. By using a stochastic geometry approach, the works \cite{han2017wirelessly} and \cite{bacha2018backscatter} studied the coverage and capacity of large-scale WPBC networks. In \cite{zhu2018inference}, an intelligent backscatter sensor system was developed, where machine learning techniques were leveraged to enhance signal processing of backscatter sensors. In \cite{lu2018wireless-powered}, a hybrid device-to-device (D2D) system integrating ambient backscattering with wirelessly powered communications was introduced, where two mode selection protocols were devised so as to adapt to diverse propagation environments. However, although power beacons benefit shortening the distance of WPT, they increase substantially the infrastructure investment of network operators \cite{Tharindu2018Simultaneous}.

Thanks to its low cost and mobility, unmanned ground vehicle (UGV) is an attractive medium to replace power beacons in WPBC for large-scale IoT networks. In \cite{Shuai2017Wirelessly}, a UGV was used to act as a mobile relay in wireless powered two-way communication system with {\it two} terminals. It is demonstrated that, with an appropriate trajectory design, a UGV can enlarge the achievable data rate region of the system. In \cite{Shuai2019Backscatter}, the optimal trajectory planning and power allocation at a UGV in a backscatter system with {\it multiple} terminals was investigated, to achieve an energy balance between UGV motion and data transmission. To the best of our knowledge, the trajectory planning of UGV and corresponding resource allocation for large-scale networks with {\it massive} terminals is still an open problem. To resolve this problem, in this paper, we design a UGV-assisted backscatter system for large-scale IoT networks, where after careful network planning, a spiral trajectory for the UGV is specified, as well as several efficient resource-allocation methods.

\subsection{Main Contributions}
This paper develops a novel network architecture where a UGV is employed to enable a large-scale backscatter network. Specifically, an AP serving as both data transmission/reception and energy carrier emitter is deployed at the center of the network while a backscatter reader is mounted on a UGV that can move along a predesigned trajectory. With the motion of UGV, the reader can visit all tags in sequence, then collect the backscattered data, and finally send them back to the AP for further processing. To fully exploit massive antennas at the AP, both half-duplex (HD) and full-duplex (FD) operation modes are studied and compared. In summary, the main contributions of this paper are as follows:
\begin{itemize}
	\item A UGV-assisted WBPC architecture suitable for large-scale IoT networks is devised. In the network, the AP transmits RF signals to activate tags. Then, the UGV moves along a predesigned trajectory and the reader mounted on the UGV collects the backscattered data. Finally, the reader relays data to the AP for further processing;	
	\item  To optimize the trajectory of UGV, the network coverage is tessellated into hexagonal cells and the optimal radius of cells is derived analytically under both FD and HD modes. Also, the optimal network planning under FD mode is proved identical to that under HD mode;	
	\item Under HD mode, the optimal resource allocation of the network is performed, including Tx/Rx beamforming at the AP and dynamic power allocation at the tag reader;
	\item Under FD mode, a successive convex approximation (SCA) based algorithm is designed to obtain the optimal power allocation at the tag reader and Tx/Rx beamforming at the AP. Moreover, two suboptimal schemes with low complexity or in closed form are developed. 
	\item Simulation results show that dynamic power allocation at the tag reader is much more energy-efficient than Tx beamforming at the AP. Also, the AP operating in FD mode has higher energy efficiency than that in HD mode.
\end{itemize}
The proposed system together with the developed algorithms can be applied, e.g., in vehicle manufacturing, where massive tags are deployed to monitor the status of various industrial equipments and UGVs are already on-site.

To detail the contributions described above, the remainder of this paper is organized as follows. Section \ref{Section-II} develops the network architecture, including network modeling and signal modeling. Section~\ref{Section-III} formulates the problem of the joint design of network planning and resource allocation, given that the AP operates in HD mode. Then, Section \ref{Section-IV} investigates the case that the AP operates in FD mode, where self-interference is accounted for. Section \ref{Section-V} presents and discusses simulation results and, finally, Section \ref{Conclusion} concludes the paper.

\textit{Notation}: Vectors and matrices are denoted by lower- and upper-case letters in boldface, respectively. Calligraphic letters indicate sets or optimization problems, depending on the context. The operators $\lfloor{x}\rfloor$ and $\lceil{x} \rceil$ represent the floor and ceiling functions of a real number $x$, respectively, while $\Re\{x\}$ returns the real part of a complex number $x$. The operators $|\bm{x}|$, $\|\bm{x}\|_2$ and $\mathbb{E}[\bm{x}]$ take the absolute value, Euclidean norm  and expectation of $\bm{x}$, respectively. The superscripts $(\cdot)^{-1}$ and $(\cdot)^{\rm H}$ mean the inverse and Hermitian transpose of a matrix, respectively.  The symbols $\mathbb{R}^{M \times L}$ and $\mathbb{C}^{M \times L}$ indicate the real and complex spaces with dimension $M \times L$, respectively. The abbreviation $x \sim \mathcal{CN}(\delta, \sigma^2)$ implies that $x$ follows a circularly symmetric complex Gaussian distribution with mean $\delta$ and variance $\sigma^2$. Finally, ${\bm{I}}$ and $\mathbb{Z}^+$ indicate an identity matrix with proper size and the set of positive integers, respectively.

\section{The Proposed Network Architecture} 
\label{Section-II} 
This section starts with network modeling, followed by the corresponding signal modeling.

\subsection{Network Modeling}
We consider a large number of backscatter tags uniformly distributed in coverage area $S$ with density $\lambda$ $\text{tags}/\text{m}^2$. An AP serving as a central processing unit or an edge computing node is located at the center of the coverage area. Each tag has no built-in battery and can only communicate with a nearby tag reader by backscattering the signals transmitted by the AP. To serve such a large-scale network, a tag reader is mounted on a UGV and shares its battery with the UGV, and the UGV moves along a carefully designed trajectory and visits all tags in succession. During each sojourn time of the UGV, the reader communicates with its nearby tags and then relays the collected data to the AP. In other words, the tag reader serves as a mobile decode-and-forward relay.  

\begin{figure}[!t]
	\centering
	\includegraphics [width=3.2in, clip, keepaspectratio]{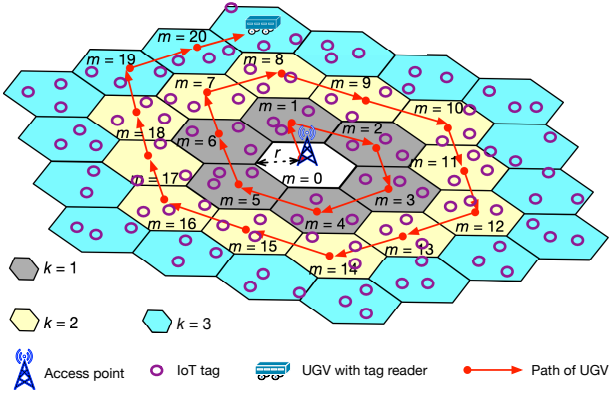}
	\caption{A UGV-assisted WPBC network.}  
	\label{Fig-1}
\end{figure}

As shown in Fig.~\ref{Fig-1}, by using a similar idea to cellular communications, we tessellate the network coverage area into hexagonal cells. Suppose that the AP is located in the central cell with ID $m = 0$, then, around the central cell, there are $M$ regular hexagonal cells and they are indexed in a spiral form with cell ID $m \in \mathcal{M} \triangleq \{1, 2, \cdots, M\}$, as shown by the red curve with arrows in Fig.~\ref{Fig-1}. For ease of notation,  $m$ is also used to denote the center of the $m^{\text{th}}$ cell as well. Moreover, all the $M$ cells are grouped into $K$ layers, indexed with $k \in \mathcal{K} \triangleq \{1, 2, \cdots, K\}$ from the inner to outer layer. For instance, the six grey cells around the central cell form the first layer (i.e., $k=1$) while the twelve yellow cells adjacent to the first layer constitute the second layer (i.e., $k=2$). Clearly, for the $k^{\rm th}$ layer, there are $6k$ cells in total. 

Due to the relatively short distance between those tags in the central cell and the AP, they are allowed to communicate directly with the AP, or, for security purposes, the central cell can be seen as an exclusion zone without any tag \cite{MXia2015TSP}. As a result, the UGV starts to move from the cell with ID $m = 1$, and then to the cell with ID $m=2$, until the last cell with ID $m = M$ in sequence (cf.  Fig.~\ref{Fig-1}). At each cell, the UGV sojourns for a while and the reader communicates with the tags in the cell and, then, relays the collected data to the AP. 

It is assumed that the reader and tags are all equipped with a single antenna while the AP has $L > 1$ antennas. The propagation channels between the reader and the AP are assumed subject to Rayleigh fading and the large-scale path-loss has a path-loss exponent $\alpha > 2$. In this section, we study the case that the AP operates in {\it half-duplex} mode, namely, all $L$ antennas at the AP are used for either data transmission or reception in time-division multiplexing. For comparison purposes, the {\it full-duplex} mode will be investigated in Section~\ref{Section-IV}. 

\begin{figure}[!t]
	\centering
	\subfigure[Frame structure of the network with AP in HD mode.]{
		\includegraphics[width=3.75in, clip, keepaspectratio]{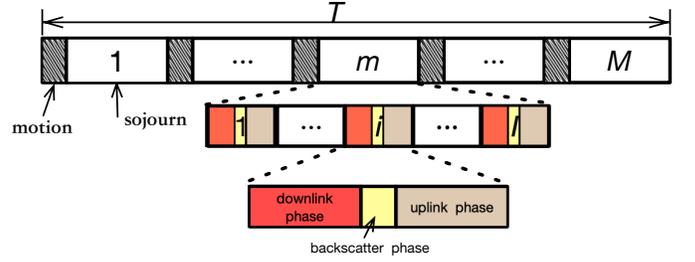} \label{Fig-2a}
	}
	\vspace{5pt}
	\subfigure[Workflow in the $(m, i)^{\rm th}$ sub-slot.]{
		\includegraphics[width=3.50in, clip, keepaspectratio]{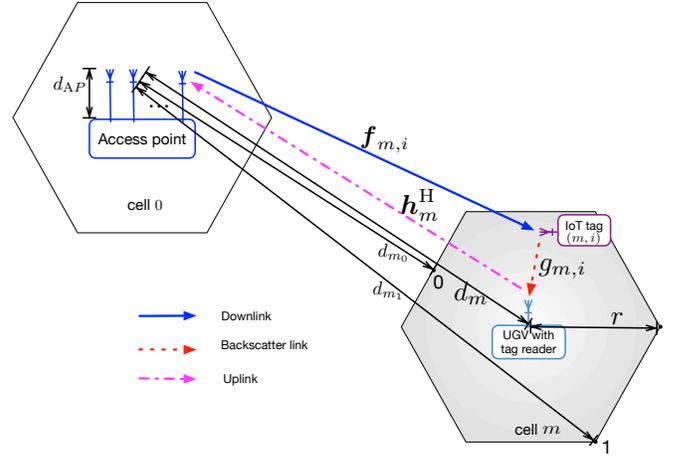} \label{Fig-2b}
	}
	\caption{Frame structure and workflow of the network.} 
	\label{Fig-2}
\end{figure}

Figure~\ref{Fig-2} illustrates the frame structure and workflow of the network shown in Fig.~\ref{Fig-1}, where the strategy of time-division multiple access (TDMA) is adopted among the tags. Specifically, as shown in Fig. \ref{Fig-2a}, a time block of length $T$ is divided into $M$ slots and the $m^{\rm th}$ slot corresponds to the sojourn duration of the UGV in the $m^{\rm th}$ cell, for all $m \in \mathcal{M}$. Each slot is further divided into $I$ sub-slots and the $i^{\rm th}$ sub-slot is devoted to the communication between the AP and the $i^{\rm th}$ tag in the cell, where $i \in \mathcal{I} \triangleq \{1, 2, \cdots, I\}$. In particular, during the $i^{\rm th}$ sub-slot when the UGV sojourns in the $m^{\rm th}$ cell, that is, in the $(m, i)^{\rm th}$ sub-slot, the communications between the AP, the $i^{\rm th}$ tag,  and the reader consists of three phases: 
\begin{itemize}
	\item {\it Downlink Phase:} The AP serving as carrier emitter broadcasts signal and the $i^{\rm th}$ tag harvests energy from its received signal;
	\item {\it Backscattering Phase:} The $i^{\rm th}$ tag modulates its data on the same RF carrier as the AP and backscatters them to the reader;
	\item {\it Uplink Phase:} The reader decodes the backscattered signal and relays the decoded data to the AP for further processing.
\end{itemize}

For a given network and UGV trajectory described above, it is clear that the energy consumed by the motion of UGV depends  on the number of cells or, equivalently, the radius of hexagonal cells. For an energy-constrained IoT network, the radius of hexagonal cells becomes the dominant factor of the network performance. Before embarking on the technical details of determining cell radius, insofar we assume that the cell radius is a prior, say, $r$, then, the area of each hexagonal cell can be given by $S_{c} = 3\sqrt{3} \, r^2/2$. Accordingly, the relationship between the coverage area $S$, the number of cells $M$, the number of cell layers $K$, and the radius $r$ can be explicitly established as 
\begin{equation} \label{Eq-1}
	M = \left\lceil \frac{S}{S_{c}}\right \rceil - 1 = \left \lceil \frac{2S}{3\sqrt{3} \, r^2}\right \rceil - 1 = \sum_{k=1}^{K}{6k} =  3K^2 + 3K.
\end{equation}
Assuming that the tags are uniformly distributed in the coverage area with density $\lambda$ $\text{tags}/\text{m}^2$, the average number of tags in each cell is
\begin{equation} \label{Eq-2}
	I = \lambda S_{c} = \frac{3\sqrt{3}}{2}\lambda r^2.
\end{equation}
By accounting for the motion time and the sojourn time of the UGV spent on all $M$ cells, the total number of sub-slots that the UGV completes visiting all tags in the network, i.e., the length of time block $T$, is explicitly computed as
\begin{equation} \label{Eq-3}
	T =  \frac{\sqrt{3} \, rM}{\nu} + MI = (3K^2 + 3K)\left(\frac{\sqrt{3} \, r}{\nu} + \frac{3\sqrt{3}}{2}\lambda r^2\right), 
\end{equation}
where $\nu$ denotes the motion speed of the UGV, in the unit of meter per sub-slot.

Now, we compute the energy consumption of the network. On the one hand, in light of \eqref{Eq-1}-\eqref{Eq-3}, the energy consumed by the UGV can be expressed as
\begin{equation} \label{Eq-4}
	E_{\rm UGV} = (\mu_1 + \mu_2\nu)\frac{\sqrt{3} \, rM}{\nu} + I\sum_{m = 1}^{M}p_{m} + TP_{\rm in}, 
\end{equation}
where $\mu_1$ and $\mu_2$ are parameters used in the mobility model of a UGV, e.g., $(\mu_1, \mu_2) =  (0.29,7.4)$ for a Pioneer 3DX robot \cite{mei2006deployment}; $p_{m}$ denotes the Tx power of tag reader for relaying data back to the AP when it sojourns at the $m^{\rm th}$ cell, and $P_{\rm in}$ refers to the power consumed by internal circuits of both the UGV and tag reader \cite{Shuai2016Multipair}. Clearly, the first term on the right-hand side (RHS) of \eqref{Eq-4} denotes the motion energy of UGV; the second term indicates the Tx energy of tag reader, and the last term accounts for the energy consumed by internal circuits of both the UGV and tag reader. On the other hand, let $\bm w_{m, i} \in \mathbb{C}^{L \times 1}$ denote the beamforming vector of AP intended for the $i^{\rm th}$ tag in the $m^{\rm th}$ cell, then, the total energy consumed by the AP can be expressed as
\begin{equation} \label{Eq-5}
	E_{\rm AP} = \sum_{m=1}^{M}\sum_{i=1}^{I}\|\bm{w}_{m, i}\|^2_2 + TP_{\rm AP},
\end{equation}
where the first term on the RHS of \eqref{Eq-5} indicates the Tx energy of AP, and $P_{\rm AP}$ in the second term stands for the power consumed by internal circuits of AP.

As illustrated in Fig.~\ref{Fig-2b}, the horizontal distance between the AP and the center of the $m^{\rm th}$ cell can be easily calculated as $\sqrt{3}\,r {k_m}$, where $k_m = \left \lceil \sqrt{12m + 6}/6 - 1/2 \right \rceil$ is the layer number in which the $m^{\rm th}$ cell is located. Given the height $d_{\rm T}$ of the AP, the Tx-Rx distance between the AP and UGV is given by 
\begin{equation} \label{Eq-6}
	d_m = \sqrt{3r^2k_m^2 + d_{\rm AP}^2} \, .
\end{equation}
Furthermore, thanks to the relatively small radius of the cells, the distance $d_{m, i}$ between the AP and the $i^{\rm th}$ tag in the $m^{\rm th}$ cell, is approximated by the distance between the AP and the center of the cell, that is, $d_{m, i} \approx d_m$, $\forall i \in \mathcal{I}$. To make the approximation as accurate as possible, a tolerable path-loss difference $\Theta$ in dB is specified as 
\begin{equation} \label{Eq-7}
	\left|10\alpha\log d_m - 10\alpha\log d_{m, i} \right| \leq \Theta.
\end{equation}
It is noteworthy that the tolerance $\Theta$ plays a pivotal role in the following network planning since the values of $d_{m, i}$ and $d_m$ depend on the cell radius $r$. 

\subsection{Signal Modeling}
As described above, in each sub-slot the communications between the AP, the $i^{\rm th}$ tag, and the reader consists of three transmission phases, namely, the downlink phase, the backscattering phase, and the uplink phase. Now, we elaborate the corresponding signal model at each phase. 

\subsubsection{Downlink Phase}
Let $s_{m ,i}$ with normalized energy (i.e., $\mathbb{E}[|s_{m ,i}|^2] = 1$) be the symbol transmitted by the AP to the $(m, i)^{\rm th}$ tag. Then, the received signal at the tag can be written as
\begin{equation} \label{Eq-8}
	x_{m, i} = d_m^{-\frac{\alpha}{2}}{\bm{f}}_{m, i}^{\rm H}{\bm{w}}_{m, i}s_{m, i} + a_{m, i},
\end{equation}
where $\bm{f}_{m, i} \in \mathbb{C}^{L \times 1}$ denotes the complex channel vector between the AP and tag $(m, i)$, and $a_{m, i} \sim \mathcal{CN}(0, \sigma_{m, i}^2)$ indicates the circular sysmmetric additive white Gaussian noise (AWGN) at tag $(m, i)$.

\subsubsection{Backscattering Phase}
When the UGV sojourns at cell $m$, the reader mounted on the UGV receives the backscattered signals from tag $(m, i)$ as well as the interfering signals from the AP. Let $c_{m,i}$ with $\mathbb{E}[|c_{m ,i}|^2] = 1$ denote the Tx symbol of tag $(m, i)$, then, under ideal synchronization, the received signal at the reader can be expressed as
\begin{align} 
	{\tilde{y}}_{m, i}\!
		&=\! g_{m, i}\sqrt{\eta} \, x_{m, i}c_{m, i} + \bm{h}_{m}^{\rm H}{\bm{w}}_{m, i} d_m^{-\frac{\alpha}{2}}s_{m, i} + \zeta_m  \label{Eq-9} \\
		&\approx\! g_{m, i}\sqrt{\eta} \, d_m^{-\frac{\alpha}{2}}\!{\bm{f}}_{m, i}^{\rm H}{\bm{w}}_{m, i}s_{m, i}c_{m, i} \!+\! d_m^{-\frac{\alpha}{2}}\!{\bm{h}}_{m}^{\rm H}{\bm{w}}_{m, i}s_{m, i} \!+\!\zeta_m, \label{Eq-10}
\end{align}
where $\eta \in [0, 1]$ denotes the power reflection coefficient of the tag; $g_{m, i}$ is the complex channel coefficient between the tag and tag reader; $\bm{h}_{m} \in \mathbb{C}^{L \times 1}$ denotes the complex channel vector between the AP and tag reader, and $\zeta_m \sim \mathcal{CN}(0, \sigma_m^2)$ is an AWGN at the reader. Notice that, the approximation in \eqref{Eq-10} is introduced by neglecting the backscattered noise $g_{m, i}\sqrt{\eta} \, a_{m, i}c_{m, i}$, which is plausible due to its relatively lower power compared to the local noise $\zeta_m$ \cite{Fuschini2008Analytical}. Moreover, due to the relatively short distance between the reader and tags compared with that between the reader and AP, the path-loss between the reader and tags is not accounted for.
	
Since the backscattered signals suffer from both downlink attenuation and backscatter attenuation, at the tag reader the strength of interfering signals from the AP is generally much stronger than that of the backscattered signals. Specifically, in light of \eqref{Eq-10}, the stronger signal, i.e., $d_m^{-\frac{\alpha}{2}}\bm{h}_{m}^{\rm H}\bm{w}_{m, i}s_{m, i}$, can be decoded firstly and then subtracted from the Rx signal ${\tilde y}_{m, i}$ \cite{bharadia2015backfi, gong2018backscatter}. Thus, the resultant signal can be written as
\begin{equation}  \label{Eq-11}
	y_{m, i} \approx g_{m, i}\sqrt{\eta} \, d_m^{-\frac{\alpha}{2}}\bm{f}_{m, i}^{\rm H}\bm{w}_{m, i}s_{m, i}c_{m, i} + \zeta_m.
\end{equation}	
Accordingly, the Rx signal-to-noise-ratio (SNR) at the reader is given by
\begin{equation}  \label{Eq-12}
	\gamma_{m, i} = \frac{\eta }{ \sigma^2_m  d_m^{\alpha}}|g_{m, i}|^2|\bm{f}_{m, i}^{\rm H}\bm{w}_{m, i}|^2.
\end{equation}

\subsubsection{Uplink Phase}
Let the Tx signal of the reader be $u_{m, i}$ with $\mathbb{E}[|u_{m, i}|^2] = 1$, which contains the decoded information from tag $(m, i)$ in the previous phase. Then, the received signal at the AP from tag $(m, i)$ is
\begin{equation}  \label{Eq-13}
	z_{m, i} =\bm{v}_{m}^{\rm H}\bm{h}_{m}d_m^{-\frac{\alpha}{2}}\sqrt{p_{m}} \, u_{m, i} + \varsigma,
\end{equation}
where $p_{m}$ and $\bm{v}_m \in \mathbb{C}^{L \times 1}$ denote the Tx power of the tag reader and the Rx beamforming vector of the AP, respectively; $\varsigma \sim \mathcal{CN}(0, \sigma^2)$ is an AWGN at the AP. As a result, the received SNR at the AP can be computed as
\begin{equation}  \label{Eq-14}
	\Gamma_{m} =  \frac{p_{m}}{\sigma^2 d_m^{\alpha} }|\bm{v}_{m}^{\rm H}\bm{h}_{m}|^2.
\end{equation}

In light of \eqref{Eq-12} and \eqref{Eq-14}, the achievable data rate between the AP and tag $(m ,i)$ can be explicitly given by
\begin{align}  \label{Eq-15}
	R_{m ,i}^{\rm HD} = &\min\left\{\frac{1}{2}\log\left(1 + \frac{\eta }{\sigma^2_md_m^{\alpha}}|g_{m, i}|^2|\bm{f}_{m, i}^{\rm H}\bm{w}_{m, i}|^2\right), \right.\nonumber\\
	& \left.\frac{1}{2}\log\left(1 + \frac{p_{m}}{\sigma^2d_m^{\alpha}}|\bm{v}_{m}^{\rm H}\bm{h}_{m}|^2\right)\right\},
\end{align}
where the factor $1/2$ before the logarithm operator is due to the dual-hop transmission, namely, from the tag to the tag reader and then to the AP.
\begin{remark}[On the CSI acquisition, control signaling, and synchronization]
In the network under study, while the tags are battery-free and cannot generate active radio signals, the channel state information (CSI) can be acquired at the tag reader through semi-blind channel estimation \cite{Zhang2019A} or blind channel estimation \cite{8320359}. As for the control signaling, following the similar procedure developed in \cite{Moeinfar2012Design}, the reader allocates an identification (ID) number to each tag, and then, if necessary, wakes up the desired one by its unique ID and communicates with it by using the RF signals from the AP. As passive tags are usually equipped with low-power oscillators, in practice, the EPC Gen2 protocol \cite{EPC2018} uses slotted Aloha to synchronize the AP and all tags in a cell. Namely, each tag adjusts its clock offset upon reception from the AP and introduces a guard time to ensure that it communicates with the reader at the allocated sub-slot \cite{Yildirim2019On}. In practice, even imperfect synchronization has little effect on the network performance \cite{Wang2012Efficient}. For more details on these engineering issues, the interested reader is referred to, e.g., \cite{Wang2012Efficient, 8737551} and references therein.
\end{remark}

\section{Joint Network Planning and Resource Allocation}
\label{Section-III}
In this section, we investigate the network planning and optimal resource allocation at the AP and tag reader, including Tx/Rx beamforming at the AP and Tx power allocation of the tag reader. As the data transmission of tag reader consumes much less energy than the motion of UGV, the objective of our design is to minimize the total energy consumption of the AP and UGV while satisfying the quality-of-service (QoS) requirement of tags and the AP. Accordingly, let $\bm{w} \triangleq \{\bm{w}_{m, i} | (m, i) \in \mathcal{M} \times \mathcal{I}\}$, $\bm{v} \triangleq \{\bm{v}_{m} | m \in \mathcal{M}\}$, and $\bm{p} \triangleq \{ p_{m} | m \in \mathcal{M}\}$, the optimization problem of joint network planning and resource allocation can be formulated as follows:
\begin{subequations}
	\begin{align}
			\mathcal{P}1:  \min_{\bm{w},\bm{v}, \bm {p}, r} \
			 & E_{\rm UGV} + E_{\rm AP}  \label{Eq-P1a} \\
			 {\rm s.t.} \
			 & R_{m ,i}^{\rm HD} \geq R_{\min}, \ \forall (m, i) \in \mathcal{M} \times \mathcal{I}, \label{Eq-P1b} \\
			 & E_{\rm UGV} \leq E_{\max}, \label{Eq-P1c}\\
			 & T \leq T_{\max}, \label{Eq-P1d}\\
			 & |10\alpha\log d_m - 10\alpha\log d_{m, i}| \leq \Theta, \nonumber\\
			  &\forall (m, i) \in \mathcal{M} \times \mathcal{I}, \label{Eq-P1e} \\
			 & 0 \leq p_{m} \leq p_{\max}, \ \forall m \in \mathcal{M}, \label{Eq-P1f}\\
			 & \|\bm{w}_{m, i}\|^2_2 \leq P_{\max}, \ \forall (m, i) \in \mathcal{M} \times \mathcal{I}, \label{Eq-P1g}\\
			 & \|\bm{v}_{m}\|^2 = 1, \  \forall m \in \mathcal{M}, \label{Eq-P1h} 
	\end{align}
\end{subequations}
where $E_{\rm UGV}$ and $E_{\rm AP}$ in \eqref{Eq-P1a} are earlier defined in \eqref{Eq-4} and \eqref{Eq-5}, respectively; $R_{\min} > 0$ in \eqref{Eq-P1b}, $E_{\max} \ge 0$ in \eqref{Eq-P1c}, and $T_{\max} > 0$ in \eqref{Eq-P1d} indicate the minimum required data rate for tag $(m, i)$, the maximum amount of energy stored in the UGV, and the maximum allowable system operating time, respectively; $p_{\max}$ in \eqref{Eq-P1f} and $P_{\max}$ in \eqref{Eq-P1g} represent the maximum Tx power of the reader and of the AP, respectively; finally, \eqref{Eq-P1h} indicates that Rx beamforming vectors are normalized in power.
 
Since \eqref{Eq-P1e} puts constraint on the cell radius (i.e., $r$), which further influences the number of cells (i.e., $M$), the number of constraints of $\mathcal{P}1$ depends heavily on the tolerable path-loss difference $\Theta$. As a result, $\mathcal{P}1$ is hard to solve, if not impossible. To tackle this difficulty, we first express $r$ in terms of $K$ using \eqref{Eq-1}: 
\begin{equation}\label{Eq-17}
	r = \left(\frac{2S}{3\sqrt{3} \, (3K^2 + 3K +1)}\right)^{\frac{1}{2}},
\end{equation}
then, inserting \eqref{Eq-17} into \eqref{Eq-3} yields 
\begin{align}
	T &= \frac{3K^2 + 3K}{\nu}\sqrt{\frac{2S}{\sqrt{3} \, (3K^2 + 3K +1)}} + {\lambda}S\frac{3K^2 + 3K}{3K^2 + 3K +1} \nonumber \\
	   &\approx \underbrace{\frac{3K^2 + 3K}{\nu}\sqrt{\frac{2S}{\sqrt{3} \, (3K^2 + 3K +1)}}}_{t(K)} + \, {\lambda}S, \label{Eq-18}
\end{align}
where $t(K)$ denotes the motion time of UGV, which is a function of the number of cell layers (i.e., $K$); and where ${\lambda}S$ stands for the sojourn time of UGV, which depends only on the number of tags in the network (by recalling the fact that time-division multiplexing is applied to tags) but is independent of $K$; and the approximation $(3K^2 + 3K)/({3K^2 + 3K +1}) \approx 1$ is introduced to reach \eqref{Eq-18}, which holds if $K \ge 2$.  Consequently, substituting \eqref{Eq-18} into \eqref{Eq-4} and \eqref{Eq-5} as well as performing some algebraic manipulations, Problem $\mathcal{P}1$ can be equivalently rewritten as
\begin{subequations} \label{Eq-19}
	\begin{align}
		\!\!\!\mathcal{P}1.1: \!\!\!\min_{\bm{w},\bm{v}, \bm{p}, K} 
		& \sum_{m=1}^{M}\sum_{i=1}^{I}\|\bm{w}_{m, i}\|^2_2 + \left(t(K) + \lambda S\right)(P_{\rm AP} + P_{\rm in}) \nonumber\\
		& \hspace{40pt} +  (\mu_1 + \mu_2\nu)t(K) + I\sum_{m = 1}^{M}p_{m} \label{P1.1-0} \\
		{\rm s.t.}\quad
		& (\mu_1 + \mu_2\nu)t(K) + I\sum_{m = 1}^{M}p_{m} + \left(t(K) + \lambda S\right)P_{\rm in}\nonumber\\
		&  \leq E_{\max}, \label{P1.1-1} \\
		& t(K) + \lambda S \leq T_{\max}, \label{P1.1-2} \\
		& K \in \mathbb{Z}^+, \label{P1.1-3}\\
		& \eqref{Eq-P1b}, \eqref{Eq-P1e}-\eqref{Eq-P1h}. \nonumber
	\end{align}
\end{subequations}

While Problem $\mathcal{P}1.1$ is still nontrivial as most variables are coupled together, after careful observation, we find that the number of cell layers (i.e., $K$) can be decoupled from others and determined separately, as detailed below.

\subsection{Optimal Number of Cell Layers}
\label{Subsection-III-A}
As illustrated in Fig.~\ref{Fig-1}, more cells introduce longer trajectory of the UGV and, thus, higher motion energy consumption. Since data transmission consumes much less energy than the motion of UGV, the energy consumed by both the UGV and tag reader is dominated by the motion of UGV, or equivalently by the number of cells. From this perspective, the number of cells should be as few as possible so as to save energy (strictly speaking, $t(K)$ is an increasing function of $K$, as proved in Appendix~\ref{Proof-Theorem-1}). On the other hand, as shown in \eqref{Eq-P1e}, the cell radius is limited by the tolerable path-loss difference. As a result, the optimal radius of hexagonal cells is the maximal value that satisfies the equality of \eqref{Eq-P1e} and the optimal number of cell layers can be determined, as summarized in the following theorem.

\begin{theorem} \label{Theorem-1}
	 The optimal number of cell layers is given by
	 \begin{equation} \label{Eq-20}
	 	K^{*} = \left\{
	 	\begin{aligned}[rl]
	     	\max\left\{1, \lceil K_B \rceil, \lceil K_I \rceil\right\}, & \text{ if } S \leq \frac{3\sqrt{3}}{2} d^2_{\rm AP}; \\
	     	K_0, & \text{ otherwise},
	     \end{aligned}\right. 
	     \end{equation}
	 where $K_B$, $K_I$, and $K_0$ are defined in \eqref{KA}, \eqref{KI}, and \eqref{K0} of Appendix~\ref{Proof-Theorem-1}, respectively.
\end{theorem}
\begin{proof}
	See Appendix~\ref{Proof-Theorem-1}.
\end{proof}

With the resultant $K^*$, the optimal radius of hexagonal cells (i.e., $r^*$), the number of cells (i.e., $M^*$), and the average number of tags in each cell (i.e., $I^*$) can be explicitly determined as
\begin{align} 
	r^* &= \left(\frac{2S}{3\sqrt{3}\left(3{K^*}^2 + 3K^* + 1\right)}\right)^{\frac{1}{2}}, \label{Eq-21a} \\
	M^* &=  3{K^*}^2 + 3K^*, \label{Eq-21b} \\
	I^* &=  \frac{S \lambda}{3{K^*}^2 + 3K^* + 1}. \label{Eq-21c}
\end{align}
Once the optimal number of layers (i.e., $K^{*}$) is determined according to Theorem~\ref{Theorem-1}, the cell radius (i.e., $r^{*}$) can be easily computed through \eqref{Eq-21a} and, finally, the spiral trajectory of the UGV is fixed as per Fig.~\ref{Fig-1}. This completes the task of optimal network planning. Next, we investigate the dynamic resource allocation.

To avoid excessive notation, in the rest of this section we remain to use $M$ and $I$, instead of $M^*$ and $I^*$ with the understanding that their optimal values are derived. After $K^*$ is determined by \eqref{Eq-20}, Problem $\mathcal{P}1.1$ can be simplified as
\begin{subequations} \label{Eq-24}
	\begin{align}
		\mathcal{P}1.2: \min_{\bm{w},\bm{v}, \bm{p}}  \
		& \sum_{m=1}^{M}\sum_{i=1}^{I}\|\bm{w}_{m, i}\|^2_2  + I\sum_{m = 1}^{M}p_{m}  \label{P2-0} \\
		{\rm s.t.} \
		& I\sum_{m = 1}^{M}p_{m} \leq C, \label{P2-1} \\
		& \eqref{Eq-P1b}, \eqref{Eq-P1f}-\eqref{Eq-P1h}, \nonumber
	\end{align}
\end{subequations}
where $C \triangleq E_{\max} - (\mu_1 + \mu_2\nu)t(K) - \left(t(K) + \lambda S\right)P_{\rm in}$ denotes the amount of energy used for data transmission. It is clear that except \eqref{Eq-P1b}, Problem $\mathcal{P}1.2$ is decoupled in $\bm{w}$ and $\{\bm{p}, \bm{v}\}$, both in the objective function and the constraints.  To break the coupling of \eqref{Eq-P1b}, on account of \eqref{Eq-15}, it is straightforward that \eqref{Eq-P1b} is equivalent to
\begin{align}
	|\bm{f}_{m, i}^{\rm H}\bm{w}_{m, i}|^2 &\geq A_{m, i}, \ \forall (m, i) \in \mathcal{M} \times \mathcal{I}, \label{Eq-25} \\
	p_{m}|\bm{v}_{m}^{\rm H}\bm{h}_{m}|^2 &\geq B_m, \ \forall m \in \mathcal{M}, \label{Eq-26}
\end{align}
where $A_{m, i} \triangleq {d_m^{\alpha}\sigma^2_m\left(4^{R_{\min}} - 1\right) }/{(\eta|g_{m, i}|^2)}$ and $B_m \triangleq d_m^{\alpha}\sigma^2\left(4^{R_{\min}} - 1\right)$. Therefore, $\mathcal{P}1.2$ can be decomposed into two independent subproblems, one for $\bm{w}$ and the other for $\{\bm{p}, \bm{v}\}$: 
\begin{align}
	\mathcal{P}1.3:  \min_{\bm{w}} \ & \sum_{m=1}^{M}\sum_{i=1}^{I}\|\bm{w}_{m, i}\|^2_2, \quad {\rm s.t.} \ \eqref{Eq-P1g}, \eqref{Eq-25}; \label{Eq-27} \\ 
	\mathcal{P}1.4:  \min_{\bm{v}, \bm{p}} \ & I\sum_{m = 1}^{M}p_{m}, \quad {\rm s.t.} \ \eqref{Eq-P1f}, \eqref{Eq-P1h}, \eqref{P2-1}, \eqref{Eq-26}. \label{Eq-28}
\end{align}

In the following, $\mathcal{P}1.3$ is first solved to obtain the Tx beamforming vector $\bm{w}$ and, then, $\mathcal{P}1.4$ is solved to obtain both the Rx beamforming vector $\bm{v}$ and the power allocation vector $\bm{p}$.

\subsection{Optimal Tx Beamforming of the AP}
Since all tags work in a time-division multiplexing fashion, $\mathcal{P}1.3$ can be further decomposed into $M \times I$ subproblems, one for each $(m, i) \in \mathcal{M}\times \mathcal{I}$:
\begin{subequations}\label{Eq-29}
\begin{align}  
	\mathcal{P}1.3':  \min_{\bm{w}_{m, i}}\
		& {\|\bm{w}_{m, i}\|^2_2}, \\
		 {\rm s.t.} \
		 &\|\bm{w}_{m, i}\|^2_2 \leq P_{\max}, \\
		&|\bm{f}_{m, i}^{\rm H}\bm{w}_{m, i}|^2 \geq A_{m, i}.
\end{align}
\end{subequations}
Thanks to the special structure of $\mathcal{P}1.3'$, $\bm{w}_{m, i}$ can be analytically derived, as formalized below.
\begin{theorem} \label{Theorem-2}
	The optimal Tx beamforming vector of the AP for the $(m, i)^{\rm th}$ tag is given by
	\begin{equation} \label{Eq-30}
		\bm{w}_{m, i}^* = \sqrt{A_{m, i}} \, \frac{\bm{f}_{m, i}}{\|\bm{f}_{m, i}\|_2^2}, \text{ if } \|\bm{f}_{m, i}\|_2 \geq \sqrt{\frac{A_{m, i}}{P_{\max}}}.
	\end{equation}
	Otherwise, Problem $\mathcal{P}1.3'$ is infeasible.
\end{theorem}
\begin{proof}
	See Appendix \ref{appendixB}.
\end{proof}

It is observed that \eqref{Eq-30} is essentially the well-known conjugate beamforming. Also, the condition in \eqref{Eq-30} implies that, if the channel quality between the AP and the $(m, i)^{\rm th}$ tag is too poor to satisfy the QoS requirement under the peak Tx power constraint of AP, $\mathcal{P}1.3'$ as well as $\mathcal{P}1$ is infeasible, and the tag is not allowed to make transmission at the moment.

\subsection{Optimal Rx Beamforming of the AP and Optimal Tx Power of the Tag Reader}
As the objective function of $\mathcal{P}1.4$ shown in \eqref{Eq-28} is monotonically increasing with $p_m \geq 0$, the term $|\bm{v}_{m}^{\rm H}\bm{h}_{m}|^2$ in \eqref{Eq-26} must be maximized so as to minimize the Tx power of tag reader (i.e., $p_m$) while satisfying \eqref{Eq-26}. Together with \eqref{Eq-P1h}, the design of $\bm{v}_{m}$ is the maximal ratio combining (MRC) problem \cite{roy2004maximal} while satisfying $\|\bm{v}_{m}\|^2 = 1$. Consequently, the optimal Rx beamforming vector of AP in slot $m$ is given by
\begin{equation} \label{Eq-31}
	\bm{v}_{m}^* = \frac{\bm{h}_{m}}{\|\bm{h}_{m}\|_2}. 
\end{equation}
With ${\bm v_{m}}$ determined by \eqref{Eq-31}, $\mathcal{P}1.4$ can be simplified as
\begin{subequations}
	\begin{align}
		\mathcal{P}1.4': \min_{\bm{p}} \
		& \sum_{m = 1}^{M}p_{m} \\
		{\rm s.t.} \
		& p_{m} \geq \frac{B_m}{|\bm{v}_{m}^{\rm H}\bm{h}_{m}|^2}, \ \forall m \in \mathcal{M}, \label{p-B} \\
		& \eqref{Eq-P1f}, \eqref{P2-1}. \nonumber
	\end{align}
\end{subequations}

Since smaller $p_{m}$ would make \eqref{Eq-P1f} and \eqref{P2-1} easier to satisfy and lead to small objective function of Problem $\mathcal{P}1.4$, it is straightforward that the optimal $p_{m}$ should satisfy the equality in \eqref{p-B}, as summarized in the following theorem:
\begin{theorem}\label{Theorem-p}
	The optimal Tx power of the tag reader in cell $m$ is given by
	\begin{equation} \label{Eq-34}
		p_{m}^* = \frac{B_m}{\|\bm{h}_{m}\|_2^2},
	\end{equation}
	if
	\begin{equation} \label{Eq-35}
		\frac{B_m}{\|\bm{h}_{m}\|_2^2} \leq p_{\max} \text{ \rm and } \sum_{m = 1}^{M}\frac{I B_m}{\|\bm{h}_{m}\|_2^2} \leq C. 
	\end{equation}
	Otherwise, Problem $\mathcal{P}1.4'$ is infeasiable.
\end{theorem}
The first condition in \eqref{Eq-35} implies that, if the channel quality between the tag reader and AP is too poor to satisfy the QoS requirement under the reader's peak Tx power constraint, the reader shall suspend its communication to the AP at the moment. The second condition in \eqref{Eq-35} means that the total energy used for data transmission in the network must be no larger than the prepared energy $C$, defined immediately after \eqref{Eq-24}. If one or both of the conditions cannot be satisfied, $\mathcal{P}1.4'$ as well as $\mathcal{P}1$ is infeasible.

\section{Full-Duplex Mode} 
\label{Section-IV}
\begin{figure}[t]
	\centering
	\subfigure[An AP with self-interference.]{
		\includegraphics[width=1.80in, clip, keepaspectratio]{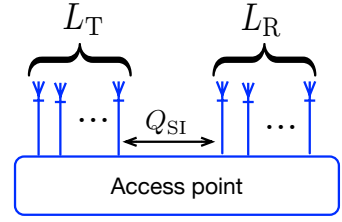} \label{Fig-3a}
	}
	\subfigure[Frame structure of the network with AP in FD mode.]{
		\includegraphics[width=3.65in, clip, keepaspectratio]{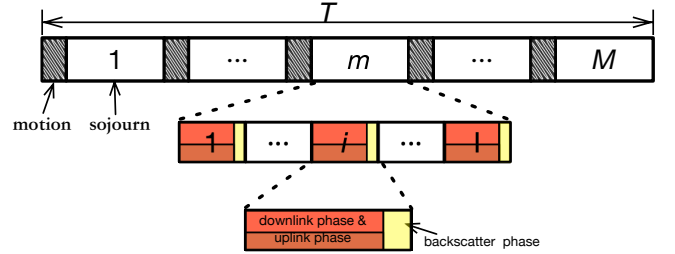} \label{Fig-3b}
	}
	\caption{An illustration of AP and frame structure operating in full-duplex mode.} 
	\label{Fig-3}
\end{figure}

Insofar we have assumed an AP with $L$ antennas operating in HD mode. Compared to HD mode, FD multiplexing achieves higher spectral efficiency and lower latency \cite{Gang2014In}. In particular, as illustrated in Fig.~\ref{Fig-3a}, the $L$ antennas at the AP are divided into two parts: $L_{\rm T}$ antennas are used for data transmission and the remaining $L_{\rm R}$ ones are for data reception, with $L_{\rm T} + L_{\rm R} = L$. As the Tx signal of the AP is generally much stronger than its Rx signal, the FD mode suffers from severe self-interference \cite{wang2020performance}. While some advanced signal processing techniques can be exploited to suppress self-interference, there is still residual self-interference. To account for this residual self-interference in our design, a self-interference channel between the Tx and Rx antennas is modeled as $\bm{Q}_{\rm SI} \in \mathbb{C}^{L_{\rm R} \times L_{\rm T}}$, which is invariant to different time slots since the antennas are relatively fixed \cite{everett2014passive, shende2017half}. In the following, we investigate the optimal network planning and resource allocation where the AP operates in FD mode while suffering from residual self-interference. Unlike the HD mode, to deal with the residual self-interference inherent in FD mode, the Rx beamforming vectors at the AP must be carefully redesigned.

Due to the self-interference, the {\it signal model} in FD mode differs significantly from the preceding HD counterpart. More specifically, in the uplink phase the Rx signal at the AP for the $(m, i)^{\rm th}$ tag is now reexpressed as
\begin{equation} \label{Eq-37}
	z_{m, i} =\bm{v}_{m, i}^{\rm H}\left(\bm{h}_{m}d_m^{-\frac{\alpha}{2}}\sqrt{p_{m}}u_{m, i} + \bm{Q}_{\rm SI}\bm{w}_{m, i}s_{m, i}\right)  + \varsigma,
\end{equation}
where the second term in the parentheses denotes the self-interference caused by the FD operation of the AP. Then, the instantaneous Rx signal-to-interference-plus-noise-ratio (SINR) at the AP can be computed as
\begin{equation} \label{Eq-38}
	\Gamma_{m, i} =  \frac{p_{m}d_m^{-\alpha}|\bm{v}_{m, i}^{\rm H}\bm{h}_{m}|^2}{|\bm{v}_{m, i}^{\rm H}\bm{Q}_{\rm SI}\bm{w}_{m, i}|^2 + \sigma^2},
\end{equation}
and accordingly, the achievable data rate between the AP and the $(m ,i)^{\rm th}$ tag is modified from \eqref{Eq-15} as
\begin{align} \label{Eq-39}
	R^{\rm FD}_{m ,i} = &\min\left\{\log\left(1 + \frac{\eta }{\sigma^2_md_m^{\alpha}}|g_{m, i}|^2|\bm{f}_{m, i}^{\rm H}\bm{w}_{m, i}|^2\right),\right. \nonumber\\
	& \left.\log\left(1 + \frac{p_{m}d_m^{-\alpha}|\bm{v}_{m, i}^{\rm H}\bm{h}_{m}|^2}{|\bm{v}_{m, i}^{\rm H}\bm{Q}_{\rm SI}\bm{w}_{m, i}|^2 + \sigma^2}\right)\right\} .
\end{align}
Therefore, the optimization problem of joint network planning and resource allocation under FD mode is the same as that in $\mathcal{P}1$, except that $R_{m, i}^{\rm HD}$ in \eqref{Eq-P1b} and $\bm{v}_m$ in \eqref{Eq-P1h} are replaced with $R^{\rm FD}_{m ,i}$ and $\bm{v}_{m, i}$, respectively.

As previously stated in Subsection~\ref{Subsection-III-A}, since the data transmission of tag reader consumes much less energy than the motion of UGV, even though the channel conditions under FD mode differ from those under HD mode, the network planning can follow the same procedure as what we have done in the preceding section. In other words, under FD mode, the radius of hexagonal cells, the number of cells, and the number of tags in each cell are also determined by \eqref{Eq-21a}, \eqref{Eq-21b}, and \eqref{Eq-21c}, respectively. Moreover, for notational simplicity, we remain to use $M$ and $I$, instead of $M^*$ and $I^*$, to denote the number of cells and average number of tags in each cell, respectively. Then, similar to the derivation from $\mathcal{P}1$ to $\mathcal{P}1.2$, the new problem introduced in this section can also be derived as $\mathcal{P}1.2$ with $R_{m,i}$ and $\bm{v}_m$ being replaced, which is now labeled as $\mathcal{P}2$.

In light of $R^{\rm FD}_{m ,i}$ given by \eqref{Eq-39}, the discontinuous constraint \eqref{Eq-P1b} is equivalent to the following two inequalities: 
\begin{align}
	 |\bm{f}_{m, i}^{\rm H}\bm{w}_{m, i}|^2 & \geq A'_{m, i}, \label{Eq-42} \\
	 \frac{p_{m}|\bm{v}_{m, i}^{\rm H}\bm{h}_{m}|^2}{|\bm{v}_{m, i}^{\rm H}\bm{Q}_{\rm SI}\bm{w}_{m, i}| ^2 + \sigma^2} & \geq B'_m, \label{Eq-43}
 \end{align}
 where $A'_{m, i} \triangleq {\sigma^2_md_m^{\alpha}\left(2^{R_{\min}} - 1\right)}/{(\eta|g_{m, i}|^2)}$ and $B'_m \triangleq d_m^{\alpha}\left(2^{R_{\min}} - 1\right)$. Although the discontinuity is resolved, the problem $\mathcal{P}2$ is still challenging to solve since the variables $\bm{w}_{m, i}$, $\bm{v}_{m, i}$, and $p_m$ are coupled all together in \eqref{Eq-43}. 

To proceed, we observe that $\bm{v}_{m, i}$ is only involved in \eqref{Eq-43}, regardless of either the objective function or other constraints of $\mathcal{P}2$. Accordingly, to minimize the transmit power $p_m$ in \eqref{Eq-43}, $\bm{v}_{m, i}$ can be determined as per
\begin{equation} \label{Eq-44}
	\bm{v}^*_{m, i} = {\rm arg}\max_{\bm{v}_{m, i}} \frac{|\bm{v}_{m, i}^{\rm H}\bm{h}_{m}|^2}{|\bm{v}_{m, i}^{\rm H}\bm{Q}_{\rm SI}\bm{w}_{m, i}| ^2 + \sigma^2}, \text{ s.t. } \eqref{Eq-P1h}.
\end{equation}
It is not hard to recognize that \eqref{Eq-44} is in the form of generalized Rayleigh quotient \cite[Sec. 4.2]{horn2012matrix}. Thus, the solution to \eqref{Eq-44} can be explicitly expressed as
\begin{equation} \label{Eq-45}
	\bm{v}^*_{m, i} = \frac{\left(\bm{Q}_{\rm SI}\bm{w}_{m, i}\bm{w}_{m, i}^{\rm H}\bm{Q}_{\rm SI}^{\rm H} + \sigma^2{\bm I}\right)^{-1}\bm{h}_{m}}{\left\|(\bm{Q}_{\rm SI}\bm{w}_{m, i}\bm{w}_{m, i}^{\rm H}\bm{Q}_{\rm SI}^{\rm H} + \sigma^2{\bm I})^{-1}\bm{h}_{m}\right\|_2}. 
\end{equation}
Inserting \eqref{Eq-45} into \eqref{Eq-43} and after performing some straightforward manipulations, \eqref{Eq-43} can be rewritten as
\begin{equation} \label{Eq-46}
	p_{m}\bm{h}_{m}^{\rm H}\left(\bm{Q}_{\rm SI}\bm{w}_{m, i}\bm{w}_{m, i}^{\rm H}\bm{Q}_{\rm SI}^{\rm H} + \sigma^2{\bm I}\right)^{-1}\bm{h}_{m} \geq B'_m.
\end{equation}
After eliminating the nontrivial matrix inversion using Sherman-Morrison formula \cite{hager1989updating}, \eqref{Eq-46} can be equivalently rewritten as:
\begin{align} \label{Eq-47}
	&\underbrace{\frac{\sigma^2}{p_{m}} B'_m \left(\|\bm{Q}_{\rm SI}\bm{w}_{m, i}\|_2^2 + \sigma^2\right)}_{W_{1}} + |\bm{h}_{m}^{\rm H}\bm{Q}_{\rm SI}\bm{w}_{m, i}|^2 \nonumber\\
	&\underbrace{- \|\bm{h}_{m}\|_2^2\|\bm{Q}_{\rm SI}\bm{w}_{m, i}\|_2^2}_{W_{2}} - \sigma^2\|\bm{h}_{m}\|_2^2 \leq 0,
\end{align}
where the term $W_{1}$ is convex (for more details, please refer to Appendix~\ref{appendixC}). Finally, the optimization problem $\mathcal{P}2$ can be reduced w.r.t. only $\bm{w}$ and $\bm{p}$:
\begin{align}
	\mathcal{P}2.1:  \min_{\bm{w},\bm{p}} \
		&\sum_{m=1}^{M}\sum_{i=1}^{I}\|\bm{w}_{m, i}\|^2_2   + I\sum_{m = 1}^{M}p_{m} \label{Eq-48} \\
		{\rm s.t.} \
		&\eqref{Eq-P1f},\eqref{Eq-P1g}, \eqref{P2-1}, \eqref{Eq-42}, \eqref{Eq-47}. \nonumber
\end{align}

The problem $\mathcal{P}2.1$ is still non-convex due to the concave constraint \eqref{Eq-42} and the concave term $W_{2}$ in \eqref{Eq-47}. In principle, by using the technique of semidefinite relaxation (SDR) \cite{Mohammadi2016Throughput, song2018joint, chalise2019full}, the non-convex $\mathcal{P}2.1$ can be transformed into a quasi-convex optimization problem, and then be solved by the bisection method \cite{boyd2004convex}. However, as the SDR-based method requires lifting the problem to higher dimensional space, it is computationally inefficient. Even worse, the SDR-based method may fail when the AP's antenna size becomes larger. To tackle the ineffectiveness of SDR-based method, we use the SCA-based method to reformulate $\mathcal{P}2.1$, as detailed below.

\subsection{SCA-based Joint Optimization (JO-SCA)}
\label{Section-IV-A}
For the convexification of \eqref{Eq-42}, the term $|\bm{f}_{m, i}^{\rm H}\bm{w}_{m, i}|^2$ is expanded as the first-order Taylor series. Then, \eqref{Eq-42} at the $n^{\rm th}$ iteration can be rewritten in terms of the optimal value of $\bm{w}_{m, i}$ at the previous iteration (i.e., $\bm{w}_{m, i}^{(n-1)}$) as
\begin{equation} \label{Eq-49}
	A'_{m, i} + \left|\bm{f}_{m, i}^{\rm H}\bm{w}_{m, i}^{(n-1)}\right|^2 - 2\Re\left\{ {\bm{w}_{m, i}^{(n-1)}}^{\rm H}\bm{f}_{m, i}\bm{f}_{m, i}^{\rm H}\bm{w}_{m, i} \right\} \leq 0, 
\end{equation}
which is linear. Likewise, \eqref{Eq-47} can be rewritten as in the following convex form:
\begin{align} \label{Eq-50}
	&\frac{\sigma^2B'_m}{p_{m}} (\|\bm{Q}_{\rm SI}\bm{w}_{m, i}\|_2^2 + \sigma^2) + |\bm{h}_{m}^{\rm H}\bm{Q}_{\rm SI}\bm{w}_{m, i}|^2 + W_{3}   \nonumber\\
	& - \sigma^2\|\bm{h}_{m}\|_2^2 \leq 0, 
\end{align}
where $W_{3} \triangleq  \|\bm{h}_{m}\|_2^2\left(\|\bm{Q}_{\rm SI}^{\rm H}\bm{w}_{m, i}^{(n-1)}\|^2  -  \right.\\ \left.2\Re\left\{\bm{w}^{\rm H}_{m, i}\bm{Q}^{\rm H}_{\rm SI}\bm{Q}_{\rm SI}\bm{w}_{m, i}^{(n-1)}\right\} \right)$ is the first-order Taylor series of the term $W_{2}$ defined in \eqref{Eq-47}. Therefore, to solve $\mathcal{P}2.1$ reduces to solve a sequence of convex problems, where at the $n^{\rm th}$ iteration the problem can be written as
\begin{align}
	\mathcal{P}2.2(n):  \min_{\bm{w},\bm{p}} \ 
	&\sum_{m=1}^{M}\sum_{i=1}^{I}\|\bm{w}^{(n)}_{m, i}\|^2_2   + I\sum_{m = 1}^{M}p^{(n)}_{m} \label{Eq-51} \\
	{\rm s.t.} \ 
	&\eqref{Eq-P1f},\eqref{Eq-P1g}, \eqref{P2-1}, \eqref{Eq-49}, \eqref{Eq-50}, \nonumber
\end{align}
which can be easily solved by using popular convex optimization package like CVX \cite{boyd2004convex}. With the optimal solutions $\bm{w}_{m, i}^{(n)*}$ to $\mathcal{P}2.2 (n)$, for all $(m, i) \in \mathcal{M} \times \mathcal{I}$, we set ${\bm w^{(n)}} = {\bm w^{(n)*}}$ and then $\mathcal{P}2.2 (n+1)$ is solved to produce ${\bm w^{(n+1)*}}$. This process iterates until it converges. In summary, based on the SCA, the overall procedure for solving the original problem $\mathcal{P}2$ is formalized in Algorithm~\ref{Algorithm1}.

\textit{Computational Complexity}: Taking CVX for instance to solve $\mathcal{P}2.2 (n)$, at each iteration, using the interior-point method yields complexity $\mathcal{O}\left(\sqrt{MIL_{\rm T}} \, \left(M^3I^3L_{\rm T}^2 + M^2I^2L_{\rm T}^3\right)\right)$ \cite[Sec. 6.6.3]{ben2001lectures}. Thus, the overall complexity of Algorithm~\ref{Algorithm1} is $\mathcal{O}\left(N_{\rm SCA}\sqrt{MIL_{\rm T}} \, \left(M^3I^3L_{\rm T}^2 + M^2I^2L_{\rm T}^3\right)\right)$, where $N_{\rm SCA}$ denotes the number of iterations.

\begin{algorithm}[!t]
	\caption{SCA-based Solution to $\mathcal{P}2$}
	\label{Algorithm1}
	\begin{algorithmic}[1]
		\REQUIRE $\mu_1$, $\mu_2$, $\nu$, $P_{\rm in}$, $P_{\rm AP}$, $\lambda$, $S$, $\eta$, $d_{\rm AP}$, $\Theta, g_{m, i}$, $\bm{f}_{m, i}$, $\bm{h}_{m}$, $\sigma_m^2$, $\sigma^2$, $R_{\rm min}$, $p_{\rm max}$, and $P_{\rm max}$;
		\STATE Compute $M^*$ and $I^*$ as per \eqref{Eq-21b} and \eqref{Eq-21c}, respectively, and then set $M = M^*$ and $I = I^*$;
		\STATE {\bf Initialization:} Set $n = 1$ and choose a feasible initial point $\bm{w}^{(0)}$;
		\REPEAT
			\STATE Solve $\mathcal{P}2.2(n)$ to obtain ${\bm w^{(n)}}$ and ${\bm p^{(n)}}$;
			\STATE $n \leftarrow n + 1$;
		\UNTIL Convegence
		\STATE $\bm{w}^* = \bm{w}^{(n)}$, $\bm{p}^* = {\bm p^{(n)}}$;
		\STATE Compute $\bm{v}_{m, i}^*$ according to \eqref{Eq-45};
		\STATE Compute $r^*$ according to \eqref{Eq-21a};
		\ENSURE $\bm{w}^*$, $\bm{v}^*$, $\bm{p}^*$, and $r^*$.
	\end{algorithmic}
\end{algorithm}

\subsection{Suboptimal Method with Equal Power Allocation (SO-EPA) at the Reader}
\label{Section-IV-B}
While the SCA-based Algorithm~\ref{Algorithm1} is more efficient than the SDR-based method, it becomes computationally challenging when the size of the problem continues to grow. For instance, considering an IoT network for vehicle manufacturing, there must be a huge number of tags or sensors for equipment and environment monitoring, asset location, and inventory control,  among others. As explicitly computed above, the computational complexity of Algorithm~\ref{Algorithm1} is proportional to $I^{3.5}$ with $I$ being the average number of tags in each cell. This high complexity becomes impractical if $I$ is large. To suit massive-access networks, in the following, we develop a suboptimal parallel algorithm whose complexity is independent of $I$. 

The main idea is to keep the Tx power of the tag reader constant in all cells, instead of dynamically adjusting it. With such an equal power allocation at the reader, we have
\begin{equation} \label{Eq-52}
	p_m = \min\left\{\frac{C}{IM}, \ p_{\max}\right\} \equiv p, \ \forall m \in \mathcal{M}. 
\end{equation}
With $p_m$ determined by \eqref{Eq-52}, $\mathcal{P}2.2(n)$ formulated in \eqref{Eq-51} reduces to 
\begin{subequations}
	\begin{align}
	\mathcal{P}3(n):  \min_{\bm{w}} \ 
	& \sum_{m=1}^{M}\sum_{i=1}^{I}\|\bm{w}_{m, i}\|^2_2 \\
	{\rm s.t.} \
	& \|\bm{w}_{m, i}\|^2_2 \leq P_{\max}, \\
	& \frac{{\sigma^2}}{p} {B'_m} \left(\|\bm{Q}_{\rm SI}\bm{w}_{m, i}\|_2^2 + \sigma^2\right) + |\bm{h}_{m}^{\rm H}\bm{Q}_{\rm SI} \bm{w}_{m, i}|^2 \nonumber\\
	& +  W_{3} - \sigma^2\| \bm{h}_{m}\|_2^2 \leq 0, \label{w-011} \\
	& \eqref{Eq-49}. \nonumber
	\end{align}
\end{subequations}
Clearly, $\mathcal{P}3 (n)$ can be decomposed into $M I$ parallel subproblems, for all $(m, i) \in \mathcal{M}\times \mathcal{I}$:
\begin{equation}  
	\mathcal{P}3.1(n):  \min_{\bm{w}_{m, i}}  \|\bm{w}_{m, i}\|^2_2 , \text{ s.t. } \eqref{w-011} \text{ and } \eqref{Eq-49}.
\end{equation}
Once the solution to $\mathcal{P}3.1(n)$ is obtained, substituting it into \eqref{Eq-45} gives the optimal Rx beamforming vector of the AP.

\textit{Computational Complexity}: As $\mathcal{P}3.1 (n)$ involves only $L/2$ variables, by using the CVX, the overall computational complexity is $\mathcal{O}\left(N_{\rm SCA}\sqrt{L/2} \, \left(L^2/4 + L^3/8\right) \right)$.

As the suboptimal method described above computes Tx/Rx beamforming vectors for different tags in parallel, the number of tags has no influence on the computational complexity. However, the computational complexity is still in proportional to $L_{\rm T}^{3.5}$ with $L_{\rm T}$ being the number of Tx antennas at the AP. Clearly, massive antennas would yield excessive burden in computation. Thus, in the following, we further develop another suboptimal method with closed-form solution. 

\subsection{Suboptimal Method with Fixed Tx Beamforming (SO-FB) at the AP} 
\label{Section-IV-C}
Now, the Tx beamforming vector at the AP is fixed and designed in a similar way as in the HD mode. Specifically, $\bm{w}_{m, i}$ is given by \eqref{Eq-30}  with $A_{m ,i}$ (defined in \eqref{Eq-25}) being replaced by $A'_{m ,i}$ (defined in \eqref{Eq-42}). Then, substituting \eqref{Eq-30} into \eqref{Eq-45} gives the optimal Rx beamforming of the AP. With $\bm{w}$ and $\bm{v}$ being fixed, the optimization of $\bm{p}$ can be formulated as
\begin{align} \label{Eq-55}
	\mathcal{P}3.2:  \min_{\bm{p}} \sum_{m = 1}^{M}p_{m}, \text{ s.t. } \eqref{Eq-P1f}, \eqref{P2-1}, \text{and } \eqref{Eq-47}.
\end{align}
The closed-form solution to $\mathcal{P}3.2$ is formalized in the following theorem.
\begin{theorem} \label{Theorem-4}
	The optimal Tx power of the tag reader in cell $m$ is $p_{m}^* = D_m$, if $0 \leq D_m \leq p_{\max}$ and $I\sum_{m = 1}^{M} D_m \leq C$, where 
\begin{align} \label{Eq-56}
	&D_m \triangleq \max\left\{\right. \nonumber\\ 
	&\left.\frac{\sigma^2B'_m\left(\|\bm{Q}_{\rm SI}\bm{w}_{m, i}\|_2^2 + \sigma^2\right)}
	{\|\bm{h}_{m}\|_2^2\|\bm{Q}_{\rm SI}\bm{w}_{m, i}\|_2^2 \!\!-\!\! |\bm{h}_{m}^{\rm H}\bm{Q}_{\rm SI}\bm{w}_{m, i}|^2 \!+ \!\sigma^2\|\bm{h}_{m}\|_2^2} \! \bigg|\! \!\ i \in \mathcal{I}, \!\eqref{Eq-30} \!\right\}\!.
\end{align}
	Otherwise, Problem $\mathcal{P}3.2$ is infeasiable.
\end{theorem}
\begin{proof}
	In light of \eqref{Eq-56}, the inequality \eqref{Eq-47} can be equivalently rewritten as $p_m \geq D_m$. Then, together with \eqref{Eq-P1f}, it is evident that $p_m$ is bounded as $D_m \leq p_m \leq p_{\max}$. By using a similar approach to deriving Theorem~\ref{Theorem-p}, if $D_m  \leq p_{\max}$, then the optimal $p_m$ that makes $\sum_{m=1}^{M}p_m$ minimal is obtained at its lower bound, yielding $p_m^* = D_m$.
\end{proof}
It is noteworthy that, while the conditions in Theorem \ref{Theorem-4} are similar to those in Theorem \ref{Theorem-p}, the problem $\mathcal{P}3.2$ being infeasible does not necessarily mean the original problem being infeasible because the solution obtained by $\mathcal{P}3.2$ is suboptimal.

\renewcommand\arraystretch{1.05}
\begin{table*}[!t]
	\caption{Simulation Parameter Setting}  
	\centering
		{\small
		\begin{tabular}{!{\vrule width1.2pt}c!{\vrule width1.2pt}c!{\vrule width1.2pt}c!{\vrule width1.2pt}}
			\Xhline{1.2pt} 
			{\bf Symbol} & {\bf Parameter} & {\bf Value} \\ 
			\Xhline{1.2pt}  
			$d_{\rm AP}$ & Height of the AP & 25 m \\ \hline
			$S$ & Coverage of the AP & 500 m$^2$ \\ \hline
			$P_{\max}$ & Maximum Tx power of the AP & 40 dBm \\ \hline
			$P_{\rm AP}$ & Power consumed by internal circuits of the AP & 0.5 W\\ \hline
			$L$ & Number of antennas of the AP & 2 - 64 \\ \hline
			$\nu$ & Velocity of the UGV & 2 m/s \\ \hline
			$E_{\max}$ & Maximum energy stored in the UGV & $10^4$ Joule \\ \hline
			$\lambda$ & Density of backscatter tags & 0.8 tag/m$^2$\\ \hline
			$\eta$& Reflection factor of backscatter tags &  0.8 \cite{Guo2018Exploiting} \\ \hline
			$p_{\max}$ & Maximum Tx power of a tag reader & 30 dBm \\ \hline
			$P_{\rm in}$ & Power consumed by internal circuits of a tag reader & 0.2 W \\ \hline
			$\Theta$ & Path-loss error tolerance & 0.4 dB \\ \hline
			$\alpha$ & Path-loss exponent & 2.8 \\ \hline
			$R_{\min}$ & Data rate requirement & 1 bps/Hz \\
			\Xhline{1.2pt}  
		\end{tabular}} \label{table}
\end{table*}

\section{Numerical Results and Discussions} 
\label{Section-V}
In this section, we present and discuss simulation results pertaining to the proposed network and developed algorithms. In the related Monte-Carlo simulation experiments, we consider a backscatter network with coverage area $500$ m$^2$, in which the tags are of density $0.8$ tag/m$^2$ \cite{bacha2018backscatter}. The AP of height 25 m is located at the center of the coverage area. The velocity of the UGV is $2$ m/s, and the parameters for the mobility model of UGV are $(\mu_1, \mu_2) = (0.29, 7.4)$ \cite{mei2006deployment}. All fading channels, including the self-interference channel at the AP operating in FD mode, i.e., the elements of $g_{m, i}, \bm{f}_{m, i}, \bm{h}_{m}$, and $\bm{Q}_{\rm SI}$, are subject to $\mathcal{CN}(0, 1)$. The noise power at the tag reader in cell $m$ is set to  $\sigma_{m}^2 = -20$ dBm, for all $m \in \mathcal{M}$, with power spectral density of $-70$ dBm/Hz over $100$ kHz bandwidth \cite{Shuai2019Backscatter}. Also, we set $L_{\rm T} = L_{\rm R}  = L/2$ for simplicity, with $L$ being an even number. For ease of retrievability, the main parameters used in our simulation experiments are summarized in Table~\ref{table}, unless specified otherwise. All simulation results reported below are obtained by making an average over $800$ Monte-Carlo trials.

It remains to mention that, while the optimization problems discussed above may be infeasible since wireless networks may suffer from poor channel quality in practice, in this section we simulate the proposed methods with feasible points for fair comparison, like \cite{li2018first}. To address the infeasibility, admission control can be applied, as discussed in \cite{zhai2013energy}. 

\subsection{Half-Duplex Mode}
\label{Section-V-A}
\begin{figure}[!t]
	\centering
	\includegraphics [width=3.5in, clip, keepaspectratio]{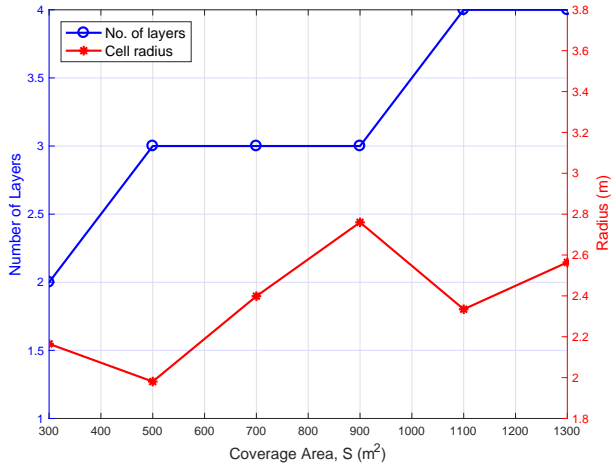}
	\caption{The effect of AP coverage area on the optimal number of cell layers and the corresponding cell radius.}
	\label{Fig-4}
\end{figure}

To illustrate the effect of network planning, Fig.~\ref{Fig-4} plots the optimal number of cell layers and the corresponding cell radius versus the AP coverage area. We observe that the optimal number of cell layers is not a monotonically increasing function but looks like a unit-step function of the coverage area. For instance, the optimal number of cell layers remains to $K^* = 3$ when the coverage area grows from $500 \ {\rm m}^2$ to $900 \ {\rm m}^2$. The reason behind this observation is that the optimization problem $\mathcal{P}1.1$ is integer programming concerning the number of cell layers. On the other hand, Fig.~\ref{Fig-4} also illustrates that the optimal cell radius does not increase monotonically with respect to the coverage area. For instance, when the coverage area grows from $300 \ {\rm m}^2$ to $500 \ {\rm m}^2$ or $900 \ {\rm m}^2$ to $1100 \ {\rm m}^2$, the cell radius decreases, because the cell radius reflects the tradeoff between the energy consumed by the motion of UGV and the energy for data transmission. This irregular cell size is why we should make optimal network planning rather than increasing the cell radius rashly. Finally, when the coverage area grows from $500 \ {\rm m}^2$ to $900 \ {\rm m}^2$, the optimal number of cell layers remains at $K^* = 3$ but the cell radius increases linearly with the coverage area since it must be seamlessly tessellated.

\begin{figure}[!t]
	\centering
	\subfigure[The effect of network planing ($L = 6$).]{
		\includegraphics[width=3.5in, clip, keepaspectratio]{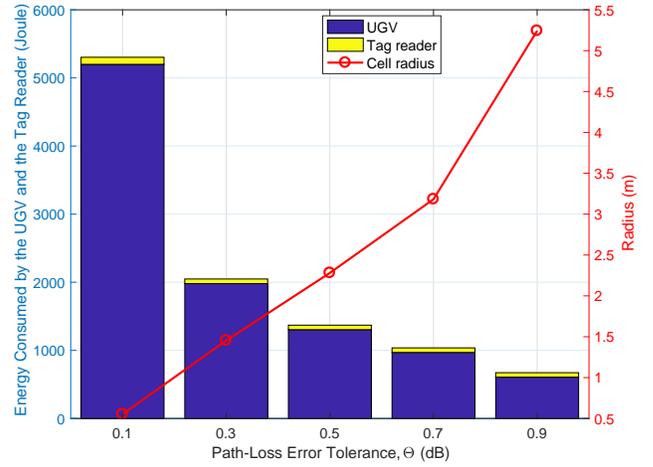}
		\label{Fig-5a}
	}
	\subfigure[The effect of multiple antennas at AP.]{
		\includegraphics[width=3.5in, clip, keepaspectratio]{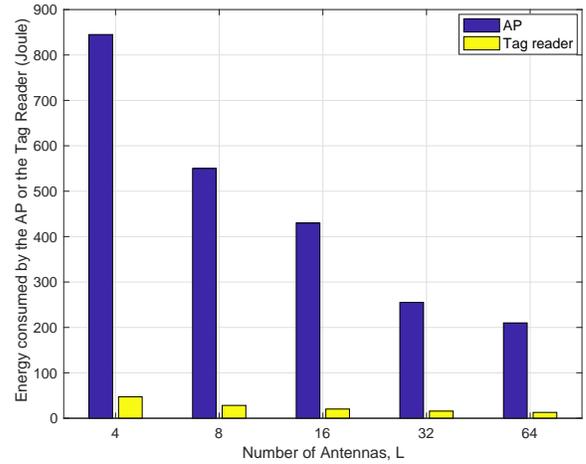}
		\label{Fig-5b}
	}
	\caption{The effects of network planing and multiple antennas on energy consumption of the network with AP in HD mode.}
\end{figure}

To show the effect of network planning on energy consumption of both the tag reader and UGV, Fig.~\ref{Fig-5a} compares the motion energy consumed by UGV with the Tx energy of tag reader versus the path-loss error tolerance defined in \eqref{Eq-7}. It is observed that the motion energy consumed by UGV decreases significantly with higher tolerance of path-loss error or, equivalently, larger radius of hexagonal cells. The reason underlying this observation is that larger cell radius implies shorter trajectory of UGV, thus consuming less motion energy. By contrast, Fig.~\ref{Fig-5a} also shows that the energy used for data transmission at tag reader is much lower than the motion energy of UGV, and it is almost unvaried with cell radius. This is because the Tx energy of tag reader is dominated by the number of tags, given a certain network coverage. On the other hand, Fig.~\ref{Fig-5a} also plots the optimal radius of hexagonal cells, which is an equivalent performance indicator of the path-loss error tolerance. By virtue of \eqref{Eq-7}, it is clear that higher tolerance of path-loss error implies larger radius of hexagonal cells, and vice versa, as illustrated by the red curve with circles in Fig.~\ref{Fig-5a}. 

To illustrate the effect of multiple antennas on energy consumption of the network, Fig.~\ref{Fig-5b} compares the energy consumed by tag reader with that by AP versus the number of antennas at AP. It is seen that both energy consumptions decrease with more antennas at AP. The reasons behind this observation are twofold: {\it i)} in the downlink phase, more antennas at AP benefits higher Tx beamforming gain to reduce Tx power, and {\it ii)} in the uplink phase, more efficient Rx beamforming at AP allows lower Tx power of tag reader. On the other hand, Fig.~\ref{Fig-5b} shows also that the energy consumed by AP is much higher than that by tag reader. This is in good accordance with the rudiments of backscatter communications, namely, as backscattered signals suffer from both downlink and backscatter attenuations, the Tx power of AP would be much higher than that of tag reader.

\subsection{Full-Duplex Mode}
\label{Section-V-B}
\begin{figure}[!t]
	\centering
	\subfigure[Comparison of different algorithms in FD mode.]{
		\includegraphics[width=3.5in, clip, keepaspectratio]{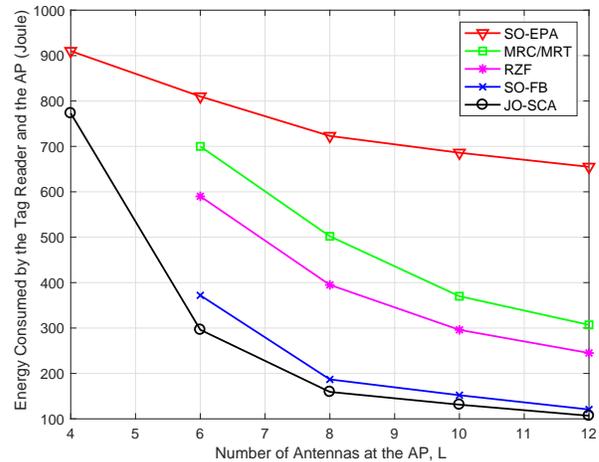}
		\label{Fig-6a}
	}
	\subfigure[HD versus FD.]{
		\includegraphics [width=3.5in, clip, keepaspectratio]{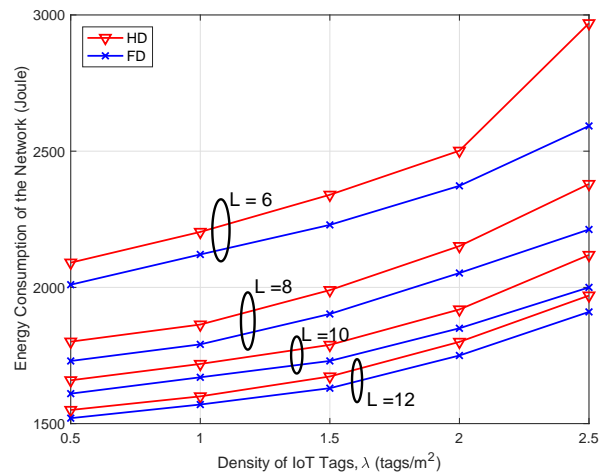}
		\label{Fig-6b}
	}
	\caption{The total energy consumption of both the tag reader and the AP.}
\end{figure}

Now, we are in a position to investigate the network performance in case the AP operates in FD mode. As discussed earlier, once the network planning is done or, equivalently, the radius of hexagonal cells is determined, the motion energy of UGV is fixed. As a result, compared with HD mode, FD mode changes only energies consumed by the AP and tag reader. Accordingly, Fig.~\ref{Fig-6a} illustrates the total energies consumed by both the AP and tag reader versus the number of antennas at the AP. In particular, three different proposed algorithms developed in Section~\ref{Section-IV} are compared, including the SCA-based joint optimization (denoted JO-SCA for short) and two suboptimal methods: one is with equal power allocation at the AP (i.e., SO-EPA), and the other is with fixed beamforming at the AP (i.e., SO-FB). For comparison purposes, two conventional methods developed in \cite{Mohammadi2016Throughput}, namely, the maximal ratio combining/maximal ratio transmission (MRC/MRT) where the Tx and Rx beamforming vectors are given by \eqref{Eq-30} and \eqref{Eq-31}, respectively, and the receive zero-forcing (RZF) where the Tx and Rx beamforming vectors are given by \eqref{Eq-30} and \eqref{Eq-45}, respectively, are also plotted in Fig.~\ref{Fig-6a} as benchmarks.

Figure~\ref{Fig-6a} illustrates that the JO-SCA algorithm consumes the least energy, at the cost of the highest computational complexity (as specified at the end of Section~\ref{Section-IV-A}). Regarding the two suboptimal algorithms, it is seen that the SO-FB consumes a slightly higher energy than the JO-SCA algorithm but the SO-EPA consumes much higher energy than the others. The rationale behind such observations is far from straightforward. Intuitively speaking, the SO-EPA algorithm requires higher computational complexity than the SO-FB algorithm, as the former needs dynamically adjust the Tx and Rx beamforming of AP whereas the Tx beamforming of the latter is fixed in closed form. However, by fixing the Tx energy of tag reader, the SO-EPA algorithm fails to make adaption to the backscattered energy obtained at the tag reader (in the downlink phase) and to the channel condition from tag reader to the AP in the uplink phase. Therefore, we conclude that, to enable lower energy consumption, dynamic power allocation at tag reader is much more efficient than Tx beamforming at the AP. On the other hand, Fig.~\ref{Fig-6a} shows that both the benchmark MRC/MRT and RZF schemes outperform the SO-EPA since they make dynamic power allocation, and the MRC/MRT underperforms the RZF because neither its Rx nor Tx beamforming is optimal. Further, the RZF scheme underperforms the SO-FB as the latter implements the transmitting zero-forcing. Therefore, the proposed SO-FB algorithm is promising in real-world applications because of its good tradeoff between energy efficiency and computational complexity.

It is noteworthy that, in our simulation experiments about Fig.~\ref{Fig-6a}, the MRC/MRT, RZF, and SO-FB algorithm are not suitable if the number of antennas at AP is less than $6$ (i.e., $L \le 6$), since the energy constraints in Theorem~\ref{Theorem-4} are not satisfied. This limit has little effect on our network design because massive MIMO  already finds wide application in practice. 

Finally, Fig.~\ref{Fig-6b} compares the energy consumption of both the tag reader and AP under HD and FD modes versus the density of tags with a different number of antennas at the AP. At first, it is seen that, whichever the HD or FD multiplexing is concerned, the total network energy consumption increases with higher tag density (i.e., $\lambda$) or fewer antennas (i.e., $L$), as expected. Then, for a fixed antenna configuration (i.e., $L$), we observe that the FD mode consumes less energy than the HD mode. The reason behind this observation is as follows. In our simulation experiments, the number of antennas at the AP remains the same whenever HD or FD mode is applied, which means that the number of Tx antennas in the FD mode is only half of that in the HD mode. Without a doubt, fewer Tx antennas will decrease the achievable data rate. Because the achievable data rate is not a linear yet a logarithmic function of the received SINR, the AP operating in FD mode with $L/2$ Tx and $L/2$ Rx antennas achieves more than half the data rate of that operating in HD mode with $L$ Tx/Rx antennas \cite[Sec. 8.2]{Tse2005Fundamentals}. Consequently, the FD mode consumes less energy than the HD mode under the same minimum data rate constraint, as illustrated in the figure. As a result, the FD mode is preferable to the HD mode in the energy-limited backscatter communications networks.

\section{Concluding Remarks} 
\label{Conclusion}
This paper developed a UGV-assisted wireless backscatter communications system for large-scale IoT networks, where the AP can operate either in half-duplex or full-duplex mode. At first, the trajectory of UGV was designed by using hexagonal tessellation, given a path-loss tolerance. Then, the optimal resource allocation, including dynamic power allocation at the tag reader and Tx/Rx beamforming at the AP, were performed. Finally, our research results disclose that, for a given antenna configuration at the AP, full-duplex multiplexing mode benefits higher energy efficiency than half-duplex mode. Also, to reduce computational complexity, fixed Tx beamforming at the AP can be allowed since it only degrades slightly the system energy efficiency, compared with the optimal resource allocation. The proposed system together with the developed algorithms can be applied, e.g., in vehicle manufacturing, where massive tags are deployed to monitor the status of various industrial equipments and UGVs are already on-site.

\begin{appendices}

\section{Proof of Theorem~\ref{Theorem-1}}
\label{Proof-Theorem-1}
	In light of the constraints of Problem $\mathcal{P}1.1$ shown in \eqref{Eq-19}, it is clear that \eqref{P1.1-1}, \eqref{P1.1-2} and \eqref{Eq-P1e} are dependent on $K$. So, we start with these constraints. Specifically, by virtue of \eqref{P1.1-1}, we have $\left(\mu_1 + \mu_2\nu + P_{\rm in}\right)t(K) \leq E_{\max} - I\sum_{m=1}^{M}p_m - \lambda SP_{\rm in} < E_{\max} - \lambda SP_{\rm in}$, which implies $t(K) < F_0 \triangleq ({E_{\max} - \lambda SP_{\rm in}})/({\mu_1 + \mu_2\nu + P_{\rm in}})$. On the other hand, the first-order derivative of $t(K)$ with respect to $K$ can be calculated as $t'(K) = (6K + 3)(\kappa + 1)\sqrt{2S/\sqrt{3}}/(2\nu\kappa\sqrt{\kappa}) > 0$ with $\kappa \triangleq 3K^2 + 3K + 1 \geq 7$, which means $t(K)$ increases with $K \geq 1$. Thus, $t(K) < F_0$ can be readily solved by means of inverse function, given by
\begin{align} \label{KF}
	& 1  \leq K  < F  \nonumber\\
		& \triangleq  \frac{1}{6}\left(9 + {\frac{3}{S}\sqrt{3}\nu^2F_0^2} + 6\nu F_0 \sqrt{\frac{{3\nu^2F_0^2}}{{4S^2}} + {\frac{2\sqrt{3}}{{S}}}}\right)^{\frac{1}{2}} \hspace{-0.75em} - \frac{1}{2}.
\end{align}
Likewise, \eqref{P1.1-2} implies that
\begin{align} \label{KG}
	& 1 \leq K \leq G   \nonumber\\
	& \triangleq \frac{1}{6}\left(9 + \frac{3}{S}\sqrt{3}\nu^2G_0^2 + 6\nu G_0 \sqrt{{\frac{3\nu^2G_0^2}{{4S^2}}} + \frac{2\sqrt{3}}{S}}\right)^\frac{1}{2} \hspace{-0.75em} - \frac{1}{2},
\end{align} 
where $G_0 \triangleq T_{\max} - \lambda S$.
	
Now, we are in a position to tackle \eqref{Eq-P1e}. We first determine the range of $d_m$. As shown in Fig.~\ref{Fig-2b} where the farthest point and the nearest point at which a tag may be located in cell $m$ from the AP are marked with ``1'' and ``0'', respectively, the distance from the AP to point ``1'' can be calculated as $d_{m_1} = \left(\left(\sqrt{3}rk_m + {\sqrt{3}r}/2\right)^2 + \left({r}/{2}\right)^2  + d_{\rm AP}^2\right)^{1/2}$, and the distance from the AP to point ``0'' is $d_{m_0} = \left(\left(\sqrt{3}rk_m - {\sqrt{3}r}/{2}\right)^2 + d_{\rm AP}^2\right)^{{1}/{2}}$. Since $d_{m_0} \leq d_{m, i} \leq d_{m_1}$ and   $d_{m_0} < d_m < d_{m_1}$, \eqref{Eq-P1e} can be decomposed into 
\begin{align}
	10\alpha\log d_{m_1} - 10\alpha\log d_{m} &\leq \Theta, \label{dm-dm0} \\ 
	10\alpha\log d_m - 10\alpha\log d_{m_0} &\leq \Theta.\label{dm-dm1} 
\end{align} 
Next, we elaborate the process to solve \eqref{dm-dm0}. \eqref{dm-dm1} can be solved in a similar way.

Substituting the expressions of $d_{m_1}$ shown above and $d_m$ given by \eqref{Eq-6} into \eqref{dm-dm0} yields
\begin{equation}\label{k-alpha}
	\frac{\left(\sqrt{3}rk_m + \frac{\sqrt{3}}{2}r\right)^2 + \left(\frac{r}{2}\right)^2  + d_{\rm AP}^2}{3r^2k^2_m + d_{\rm AP}^2} \leq 10^{\frac{\Theta}{5\alpha}}. 
\end{equation}
Relaxing $k_m$ to a continuous variable, say, $\tau$ with $1 \leq \tau \leq K$, \eqref{k-alpha} can be rewritten as
\begin{equation} \label{f1}
	\underbrace{\frac{\left(\sqrt{3}r\tau + \frac{\sqrt{3}}{2}r\right)^2 + \left(\frac{r}{2}\right)^2  + d_{\rm AP}^2}{3(r\tau)^2 + d_{\rm AP}^2}}_{f_1(\tau)} \leq 10^{\frac{\Theta}{5\alpha}}.
\end{equation}
As illustrated in Fig.~\ref{Fig-7}(a), $f_1(\tau)$ defined in \eqref{f1} is a convex quadratic function, and its maximizer can be calculated as
\begin{equation}\label{xi}
	\xi = \!\sqrt{\frac{1}{9} + \frac{d^2_{\rm AP}}{3r^2}} - \frac{1}{3} = \left(\frac{1}{9} + \frac{d^2_{\rm AP}}{2S}\sqrt{3}(3K^2+3K+1)\right)^{\frac{1}{2}} - \!\frac{1}{3}.
\end{equation}
By recalling $1 \le \tau \le K$, to compare $\xi$ with $K$, an auxiliary function is constructed as 
\begin{align} \label{f2c}
	f_2(K) &\triangleq \left(\xi + \frac{1}{3}\right)^2 -  \left(K + \frac{1}{3}\right)^2 \nonumber\\
	&= \left(\frac{3\sqrt{3}d^2_{\rm AP}}{2S} \!- \!1\right)K^2 \!+\! \left(\frac{3\sqrt{3}d^2_{\rm AP}}{2S} \!-\! \frac{2}{3}\right)\!K \!+\! \frac{\sqrt{3}d^2_{\rm AP}}{2S}.
\end{align}
As shown in Fig.~\ref{Fig-7}(b) and Fig.~\ref{Fig-7}(c), if $S \leq 3\sqrt{3}d^2_{\rm AP}/2$ or $S > 3\sqrt{3}d^2_{\rm AP}/2$ together with $K \leq K_A \triangleq \left({9\sqrt{3}d^2_{\rm AP} - 4S + \sqrt{16 - 81d_{\rm AP}^4/S^2} }\right)/\left({12S - 18\sqrt{3}d^2_{\rm AP}}\right)$, $f_2(K) > 0$, which means $K \in (0, \xi]$. On the other hand, if $S > 3\sqrt{3}d^2_{\rm AP}/2$ together with $K > K_A$, $f_2(K) < 0$, which gives $K \in (\xi, \infty)$. Therefore, \eqref{f1} can be discussed with the following two cases:

\begin{figure}[!t]
	\centering
	\includegraphics [width=3.6in, clip, keepaspectratio]{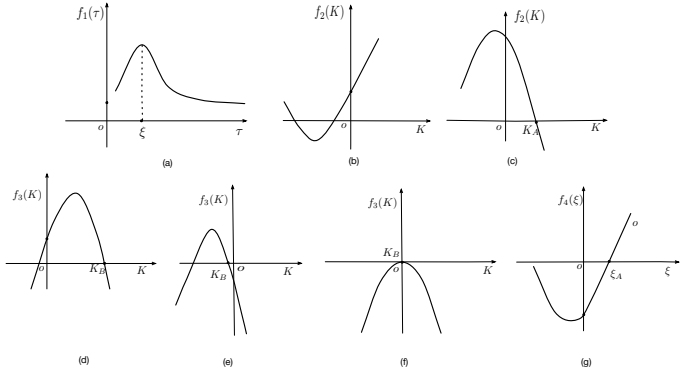}
	\caption{(a) $f_1(\tau)$; (b) $f_2(K)$ with $S \leq 3\sqrt{3}d^2_{\rm AP}/2$; (c) $f_2(K)$ with $S > 3\sqrt{3}d^2_{\rm AP}/2$; (d) $f_3(K)$ with $S > 3\sqrt{3}d^2_{\rm AP}\theta/2$; (e) $f_3(K)$ with $S < 3\sqrt{3}d^2_{\rm AP}\theta/2$; (f) $f_3(K)$ with $S = 3\sqrt{3}d^2_{\rm AP}\theta/2$; (g) $f_4(\xi)$.}
	\label{Fig-7}
\end{figure}

{\it Case I}: $S \leq 3\sqrt{3}d^2_{\rm AP}/2$ or $S > 3\sqrt{3}d^2_{\rm AP}/2$ together with $K \leq K_A$. In this case, the maximum of $f_1(\tau)$ is $f_1(K)$. Thus, \eqref{f1} can be rewritten as $f_1(K) \leq 10^{\frac{\Theta}{5\alpha}}$, which reduces to
\begin{align} \label{f3K}
	f_3(K) \triangleq& -\left(2S + 3\sqrt{3}d^2_{\rm AP}\right)\theta K^2 + \left(2S - 3\sqrt{3}d^2_{\rm AP}\theta\right)K \nonumber\\
	&+ \frac{2S}{3} - \sqrt{3}d^2_{\rm AP}\theta \leq 0,
\end{align}
where $\theta \triangleq 10^{\frac{\Theta}{5\alpha}} - 1$. Then, \eqref{f3K} can be dealt with two sub-cases: If $S > 3\sqrt{3}d^2_{\rm AP}\theta/2$, as shown in Fig.~\ref{Fig-7}(d), $K$ is bounded as
\begin{equation} \label{KA}
	K \geq K_B \triangleq  \frac{2S - 3\sqrt{3}d^2_{\rm AP}\theta+ \sqrt{\Delta_a}}{2\theta\left(2S + 3\sqrt{3}d^2_{\rm AP}\right)},
\end{equation}
with $\Delta_a \triangleq \left(2S - 3\sqrt{3}d^2_{\rm AP}\theta\right)^2 + 4\theta\left(2S + 3\sqrt{3}d^2_{\rm AP}\theta\right)\left(2S/3 - \sqrt{3}d^2_{\rm AP}\theta \right)$. On the other hand, if $S \leq 3\sqrt{3}d^2_{\rm AP}\theta/2$, as shown in Figs.~\ref{Fig-7}(e) and \ref{Fig-7}(f), \eqref{f3K} is always satisfied with $K \geq 1$.
			
{\it Case II}: $S > 3\sqrt{3}d^2_{\rm AP}/2$ together with $K > K_A$. Now, the maximum of $f_1(\tau)$ is $f_1(\xi)$. Thus, \eqref{f1} can be rewritten as $f_1(\xi) \leq 10^{\frac{\Theta}{5\alpha}}$, which implies $f_4(\xi) \triangleq 108d_{\rm AP}^4\theta^2\xi^2 - 2\sqrt{3}Sd_{\rm AP}^2(3 + 2\theta)^2\xi - 4S^2(3 + 4\theta)/3 \geq 0$. As shown in Fig.~\ref{Fig-7}(g), we obtain $\xi \geq \xi_A$, where
\begin{equation} \label{Eq-xi}
	\xi_A \!\triangleq\! \frac{{2\sqrt{3}S(2\theta + 3)^2 \!+\! \left(12S^2(2\theta + 3)^4 + 576S^2\theta^2(4\theta + 3)\right)^\frac{1}{2}}}{216d_{\rm AP}^2\theta^2}.
\end{equation}
Substituting \eqref{xi} into \eqref{Eq-xi} gives
\begin{equation}
	K \geq \frac{\sqrt{12\xi_A - 3}}{6} - \frac{1}{2}.
\end{equation}
	
Combining the preceeding \textit{Case I} and \textit{Case II}, we infer that $K$ is bounded as
	\begin{equation}\label{KAB}
		\left\{\begin{array}{l}
			K \geq  K_B,  \text{ if } S \leq \frac{3\sqrt{3}d^2_{\rm AP}}{2}; \\
			K_B \leq K \leq K_A \text{ or } K \geq \\
			\hspace{30pt}\max \left\{K_A, \frac{1}{6}\sqrt{12\xi_A - 3} - \frac{1}{2}\right\}, \text{ otherwise}.
		\end{array}\right.
	\end{equation}
By using a similar approach as above, \eqref{dm-dm1} implies that
\begin{equation} \label{K12}
	\left\{\begin{array}{l}
		K \geq  K_{I}, \text{ if } S \leq \frac{3\sqrt{3}d^2_{\rm AP}}{2}; \\
		K_{I} \leq K \leq K_{II} \text{ or } K \geq \\
		\hspace{30pt}\max\left\{K_{II}, \frac{1}{6}\sqrt{12\xi_{I} - 3} - \frac{1}{2}\right\}, \text{ otherwise},
	\end{array}\right.
\end{equation}
where
\begin{equation}\label{KI}
	K_{I} \triangleq \frac{{2S10^{\frac{-\theta}{5\alpha}} - 3\sqrt{3}d^2_{\rm AP}\theta + \sqrt{\Delta_1}}}{{2\theta\left(2S + 3\sqrt{3}d^2_{\rm AP}\right) }},
\end{equation}
with $\Delta_1 = \left(2S10^{\frac{-\theta}{5\alpha}} - 3\sqrt{3}d^2_{\rm AP}\theta\right)^2 - 4\theta\left(\sqrt{3}d^2_{\rm AP}\theta + 10^{\frac{-\theta}{5\alpha}}S/2\right)\left(2S + 3\sqrt{3}d^2_{\rm AP}\right)$, 
$\xi_{I} \triangleq\\ \left({\sqrt{3}S(1-2\theta')}\right) /\left({24d^2_{\rm AP}\theta'}\right)$, 
$\theta' \triangleq \left(1-10^{\frac{\Theta}{5\alpha}}\right)^2/\left(1+10^{\frac{\Theta}{5\alpha}}\right)^2$, 
and $K_{II} \triangleq \left(3\sqrt{3}d^2_{\rm AP} \!+\! S \!+ \! \left(14\sqrt{3}Sd_{\rm AP}^2 \!+ \!S^2 \!-\! 9d_{\rm AP}^4\right)\right)\!/\!\left({4S \!-\! 6\sqrt{3}d^2_{\rm AP}}\right)$.
	
In summary, integrating \eqref{KF}, \eqref{KG}, \eqref{KAB}, \eqref{K12} and \eqref{P1.1-3} the domain of $K$ is
\begin{equation} \label{bound}
	\mathcal{D} = 
	\left\{\begin{aligned}[rl]
		\left[\max \{ K_B ,  K_{I} \}, \min\{F ,  G\}\right] \cap \mathbb{Z}^+, & \text{ if } S \leq \frac{3\sqrt{3}d^2_{\rm AP}}{2}; \\
		\mathcal{D}_a \cap \mathcal{D}_i \cap \mathcal{D}_f \cap \mathbb{Z}^+, & \text{ otherwise},
	\end{aligned}\right.
\end{equation}
where 
$\mathcal{D}_a \triangleq \left[K_B, K_A]  \cup [\max\left\{K_A, \sqrt{12\xi_A - 3}/6 - 1/2\right\}, \right.\\ \left.+\infty\right]$, 
$\mathcal{D}_f \triangleq [1, \min\{F ,  G\}]$, 
and $\mathcal{D}_i \triangleq \left[K_{I}, K_{II}] \cup [\max\left\{K_{II}, \sqrt{12\xi_{I} - 3}/6 - 1/2\right\}, +\infty\right]$.

As the cell radius $r$ or the cell layers $K$ has little effect on channel conditions in the considered network, the objective function \eqref{P1.1-0} of Problem $\mathcal{P}1.1$ increases with $t(K)$, which is monotonically increasing with $K \geq 1$. Consequently, to minimize this objective function, $K$ must reach the lower bound of \eqref{bound}. Therefore, if $S \leq 3\sqrt{3}d^2_{\rm AP}/2$, then we obtain
\begin{equation}
	K^* = \max\left\{1, \lceil K_B \rceil, \lceil K_I \rceil\right\}.
\end{equation}
Otherwise, $K^* = K_0$, where
\begin{equation} \label{K0}
	K_0 \triangleq \left\{\begin{array}{ll}
		\max\left\{1, \lceil K_B \rceil, \lceil K_I \rceil\right\},  \text{ if } \mathcal{S}_0 \cap \mathcal{S}_I \neq \emptyset; \\
		\max\left\{1, \lceil K_I \rceil, \lceil\frac{1}{6}\sqrt{12\xi_A - 3} - \frac{1}{2} \rceil\right\}, \\
		\hspace{60pt} \text{ if }  \mathcal{S}_0 \cap \mathcal{S}_I = \emptyset \ \text{and}\ \mathcal{S}_I \cap \mathcal{S}_{X} \neq \emptyset; \\
		\max\left\{1, \lceil K_B \rceil, \lceil\frac{1}{6}\sqrt{12\xi_1 - 3} - \frac{1}{2} \rceil\right\}, \\
		\hspace{60pt}\text{ if } \mathcal{S}_0 \cap \mathcal{S}_I = \emptyset \ \text{and}\ \mathcal{S}_0 \cap \mathcal{S}_{Y} \neq \emptyset; \\
		\max\left\{\lceil K_B \rceil, \lceil K_{II},  \rceil\lceil\frac{1}{6}\sqrt{12\xi_A - 3} - \frac{1}{2},\right.\\
		 \hspace{65pt}\left.\lceil\frac{1}{6}\sqrt{12\xi_1 - 3} - \frac{1}{2} \rceil\right\},  \text{ otherwise}, \end{array} \right.
\end{equation}
with $\mathcal{S}_0 \triangleq [\max\left\{1, \lceil K_B \rceil \right\}, \lfloor K_A \rfloor]$, $\mathcal{S}_I \triangleq [\max\left\{1, \lceil K_I \rceil \right\}, \lfloor K_{II} \rfloor]$, $\mathcal{S}_{X} \triangleq [\lceil\sqrt{12\xi_A - 3}/6 - 1/2\rceil, \infty]$, and $\mathcal{S}_{Y} \triangleq [\lceil\sqrt{12\xi_I - 3}/6 - 1/2\rceil, \infty]$. It is noteworthy that the four cases in \eqref{K0} are arranged in a set manner. For instance, the first line means if the sets $[K_B, K_A], [K_I, K_{II}]$ are nonempty and intersect with each other, the lower bound of the second line in \eqref{bound} is obviously $\max\left\{1, \lceil K_B \rceil, \lceil K_I \rceil\right\}$. Finally, the results derived above are formalized in Theorem~\ref{Theorem-1}.
	
\section{Proof of Theorem~\ref{Theorem-2}}
\label{appendixB}
	By using the maximal ratio transmitting (MRT) strategy, $\bm{w}_{m, i}$ can be determined as
	\begin{equation}\label{w-P}
		\bm{w}_{m, i} = \sqrt{ P_{m, i}} \ \frac{\bm{f}_{m, i}}{\|\bm{f}_{m, i}\|_2},
	\end{equation}
	where $P_{m, i}$ represents the Tx power of the AP. Thus, obtaining the optimal $\bm{w}_{m, i}$ of Problem $\mathcal{P}1.3'$ is equivalent to obtaining the optimal $P_{m, i}$ of the following problem:
	\begin{subequations}
		\begin{align}
		\mathcal{P}(P_{m, i}):  
		& \min_{P_{m, i}} 
		P_{m, i}\\
		{\rm s.t.} \
		& P_{m, i} \leq P_{\max}, \label{P-b}\\ 
		& \sqrt{ P_{m, i}}\bm{f}_{m, i}^{\rm H}\frac{\bm{f}_{m, i}}{\|\bm{f}_{m, i}\|_2} \geq \sqrt{A_{m, i}}\label{P-c}.
		\end{align}
	\end{subequations}
	According to \eqref{P-b} and \eqref{P-c}, $P_{m, i}$ is bounded as $A_{m, i}/{\|\bm{f}_{m, i}\|^2_2 \leq	P_{m, i} \leq P_{\max}}$. Since the objective function is monotonically increasing, it is evident that if
	\begin{equation}\label{f-P}
		\frac{A_{m, i}}{\|\bm{f}_{m, i}\|^2_2} \leq P_{\max},
	\end{equation}
	the optimal $P_{m, i}$ of Problem $\mathcal{P}(P_{m, i})$ is obtained at its lower bound as
	\begin{equation}\label{P-f}
		P_{m, i}^* = \frac{A_{m, i}}{\|\bm{f}_{m, i}\|^2_2} . 
	\end{equation}
	However, if \eqref{f-P} does not hold, we cannot obtain $P_{m, i}$ under the constraints, thus, the initial problem is infeasible. Finally, substituting \eqref{P-f} into \eqref{w-P} and rearranging \eqref{f-P} complete the proof.

\section{Derivation of \eqref{Eq-47}}
\label{appendixC}
	Define ${\bm \Delta} \triangleq ({\bm r}{\bm s}^{\rm H} + {\bm O})^{-1}$ with ${\bm r} = {\bm s} = \bm{Q}_{\rm SI}\bm{w}_{m, i}$ and ${\bm O} = \sigma^2{\bm I}$, then, the inequality \eqref{Eq-46} can be rewritten as
	\begin{equation}\label{p-D-b}
		p_{m}\bm{h}_{m}^{\rm H}{\bm \Delta}\bm{h}_{m} \geq B'_m.
	\end{equation}
	With the help of Sherman-Morrison formula \cite{hager1989updating}, ${\bm \Delta}$ can be equivalently expressed as
	\begin{equation}
		{\bm \Delta} \! = {\bm {O}}^{-1} - \frac{{\bm O}^{-1}{\bm r}{\bm s}^{\rm H}{\bm O}^{-1} }{1 + {\bm s}^{\rm H}{\bm O}^{-1} {\bm r}} \!= \frac{\bm I}{\sigma^2} - \frac{\bm{Q}_{\rm SI}\bm{w}_{m, i}\bm{w}_{m, i}^{\rm H}\bm{Q}_{\rm SI}^{\rm H}}{\sigma^2\bm{w}_{m, i}^{\rm H}\bm{Q}_{\rm SI}^{\rm H}\bm{Q}_{\rm SI}\bm{w}_{m, i} + \sigma^4}. \label{D-Q}
	\end{equation}
	Substituting \eqref{D-Q} into \eqref{p-D-b} yields
	\begin{equation}\label{pB1}
		\frac{p_{m}}{\sigma^2} \left(\|\bm{h}_{m}\|_2^2 - \frac{|\bm{h}_{m}^{\rm H}\bm{Q}_{\rm SI}\bm{w}_{m, i}|^2}{\|\bm{Q}_{\rm SI}\bm{w}_{m, i}\|_2^2+ \sigma^2}\right)  \geq B'_m,
	\end{equation}
	which in turn gives \eqref{Eq-47}. Moreover, as for $W_{1}$ defined in \eqref{Eq-47}, it is clear that $\|\bm{Q}_{\rm SI}\bm{w}_{m, i}\|_2^2/p_{m}$ is jointly convex w.r.t. $\bm{w}_{m, i}$ and $p_{m} > 0$ (see the proof in \cite[Sec. 3.2.6]{boyd2004convex}), and that $\sigma^2/p_m$ is convex w.r.t. $p_m > 0$ as well. Therefore, $W_{1}$ is jointly convex w.r.t. $\bm{w}_{m, i}$ and $p_{m}$.
\end{appendices}

\bibliographystyle{IEEEtran}
\bibliography{references_EC}

\begin{IEEEbiography}[{\includegraphics[width=1in, height=1.25in, clip, keepaspectratio]{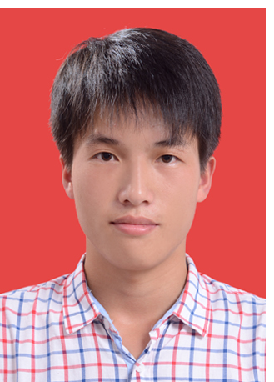}}]{Erhu Chen} received the B.E. degree in telecommunication engineering from Hubei Normal University, Huangshi, China, in 2015 and the M.S. degree in electronics and communication engineering from Sun Yat-sen University, Guangzhou, China, in 2017. He is currently working towards the Ph.D. degree in School of Electronics and Information Technology, Sun Yat-sen University, Guangzhou, China. His research interests are wireless power transfer/energy harvesting and the Internet of Things.
\end{IEEEbiography}

\begin{IEEEbiography}[{\includegraphics[width=1in, height=1.25in, clip, keepaspectratio]{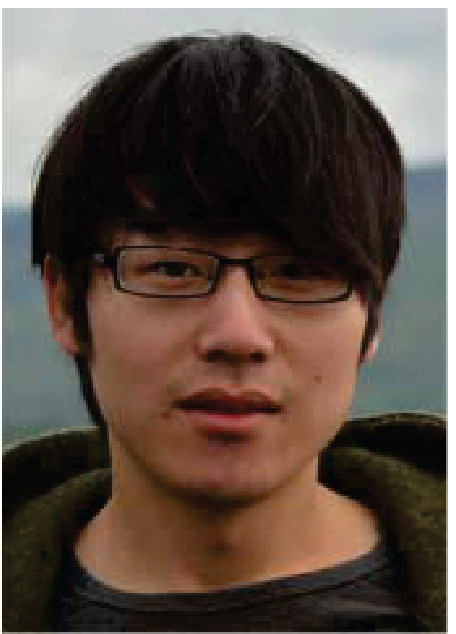}}]{Peiran Wu}(Member, IEEE) received the Ph.D. degree in electrical and computer engineering at the University
	of British Columbia (UBC), Vancouver, Canada, in 2015.
	
	From October 2015 to December 2016, he was a Postdoctoral Fellow at the same university. In summer 2014, he was a Visiting Scholar at the Institute for Digital Communications, Friedrich-Alexander-University Erlangen-Nuremberg (FAU),	Erlangen, Germany. Since February 2017, he has been with the Sun Yat-sen University, Guangzhou, China, where he is now an Associate Professor. Since 2019, he has been an Adjunct Associate Professor with the Southern Marine Science and Engineering Guangdong Laboratory, Zhuhai, China. His research interests include mobile edge computing, wireless power transfer, and energy-efficient wireless communications.
	
	He was the recipient of the Fourth-Year Fellowship in 2010, the C. L. Wang Memorial Fellowship in 2011, Graduate Support Initiative (GSI) Award in 2014 from the UBC, German Academic Exchange Service (DAAD) Scholarship in 2014, and the Chinese Government Award for Outstanding Self-Financed Students Abroad in 2014.
\end{IEEEbiography}

\begin{IEEEbiography}[{\includegraphics[width=1in, height=1.25in, clip, keepaspectratio]{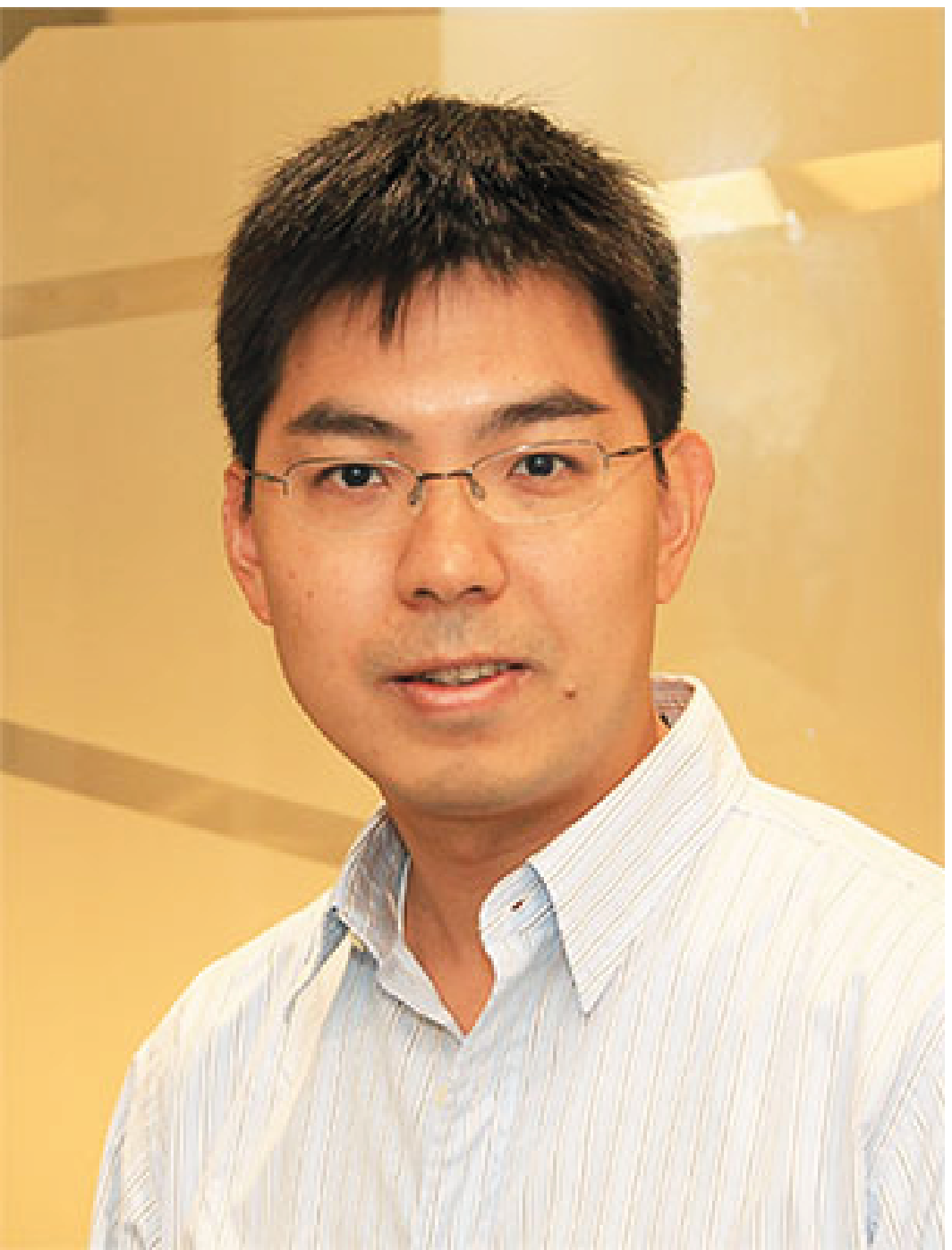}}]{Yik-Chung Wu}(Senior Member, IEEE) received the B.Eng. (EEE) degree in 1998 and the M.Phil. degree in 2001 from the University of Hong Kong (HKU),	and Ph.D. degree from Texas A\&M University in 2005.
	
	From 2005 to 2006, he was with the Thomson	Corporate Research, Princeton, NJ, USA, as a Member of	Technical Staff. Since 2006, he has been with HKU, currently as an Associate Professor. He was a visiting scholar at	Princeton University, in summers of 2015 and 2017.	His research interests are in general areas of signal	processing, machine learning and communication systems.
	
	Dr. Wu served as an editor for {\scshape IEEE Communications Letters} and {\scshape IEEE Transactions on Communications}.  He is currently an editor of {\scshape Journal of Communications and Networks}.
\end{IEEEbiography}

\begin{IEEEbiography}[{\includegraphics[width=1in, height=1.25in, clip, keepaspectratio]{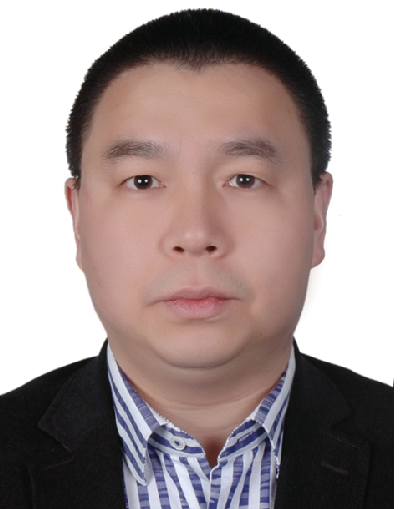}}]{Minghua Xia} (Senior Member, IEEE) received the Ph.D. degree in telecommunications and information	systems from Sun Yat-sen University, Guangzhou, China, in 2007.
	
	From 2007 to 2009, he was with the Electronics and Telecommunications Research Institute (ETRI), South Korea, and with Beijing Research and Development Center, Beijing, China, where he worked as a member and then as	a senior member of engineering staff. From 2010	to 2014, he was in sequence with The University of Hong Kong, Hong Kong, China; King Abdullah University of Science and Technology, Jeddah, Saudi Arabia; and the Institut National de la Recherche Scientifique (INRS),	University of Quebec, Montreal, Canada, as a Post-doctoral Fellow. Since	2015, he has been a Professor with Sun Yat-sen University. Since 2019, he has also been an Adjunct Professor with the Southern Marine Science and	Engineering Guangdong Laboratory (Zhuhai). His research interests are in the	general areas of wireless communications and signal processing.
\end{IEEEbiography}

\end{document}